\documentclass[aps,pra,reprint,twocolumn,showpacs,showkeys,address,bibliography]{revtex4-2}
\usepackage{amsmath,amssymb,amstext}
\usepackage[usenames,dvipsnames]{color}
\usepackage{graphicx}
\usepackage[english]{babel}
\usepackage{wasysym}
\usepackage{bm,bbold,braket}
\usepackage{natbib}
\usepackage{txfonts, comment, stmaryrd}
\usepackage{dcolumn}
\usepackage{graphicx}
\usepackage{subfigure}
\usepackage{float}
\usepackage[dvipsnames]{xcolor}
\usepackage[colorlinks,bookmarks=false,citecolor=blue,linkcolor=red,urlcolor=blue]{hyperref}
\begin{document}
\title{Faraday patterns, spin textures, spin-spin correlations and competing instabilities in a driven spin-1 antiferromagnetic Bose-Einstein condensate}
\author{Vaishakh Kargudri}
   \thanks{These authors contributed equally to this work.}
      \affiliation{Department of Physics, Indian Institute of Science Education and Research, Pune 411 008, Maharashtra, India}                      
\author{Sandra M. Jose}
   \thanks{These authors contributed equally to this work.}
      \affiliation{Department of Physics, Indian Institute of Science Education and Research, Pune 411 008, Maharashtra, India}                        
           \author{Rejish Nath}
       \affiliation{Department of Physics, Indian Institute of Science Education and Research, Pune 411 008, Maharashtra, India}
\date{\today}

\begin{abstract}
We study the formation of transient Faraday patterns and spin textures in driven quasi-one-dimensional and quasi-two-dimensional spin-1 Bose-Einstein condensates under the periodic modulation of $s$-wave scattering lengths $a_0$ and $a_2$, starting from the anti-ferromagnetic phase. This phase is characterized by a Bogoliubov spectrum consisting of three modes: one mode is gapped, while the other two are gapless. When $a_0$ is modulated and half of the modulation frequency lies below the gapped mode, density and spin Faraday patterns emerge. In that case, in quasi-one-dimension, the spin texture is characterized by periodic domains of opposite $z$-polarizations. When driven above the gap, the spin texture is characterized by random orientations of spin vectors along the condensate axis. Qualitatively new features appear in the driven quasi-two-dimensional condensate. For instance, when driven above the gap, the spin textures are characterized by anomalous vortices and antivortices that do not exhibit phase winding in individual magnetic components. Below the gap, the spin texture exhibits irregular ferromagnetic patches with opposite polarizations. The spatial spin-spin correlations in quasi-one-dimension exhibit a Gaussian envelope, whereas they possess a Bessel function dependence in quasi-two-dimension. Under the $a_2$-modulation, the density patterns dominate irrespective of the driving frequency, unless the spin-dependent interaction strength is sufficiently smaller than that of the spin-independent interaction. The intriguing scenario of competing instability can emerge when both scattering lengths are simultaneously modulated. Finally, we show that the competing instabilities result in a complex relationship between the population transfer and the strength of the quadratic Zeeman field, while keeping all other parameters constant.
\end{abstract}
\maketitle

\section{Introduction}
Periodically driven spinor Bose-Einstein condensates (BECs) offer a unique platform to explore fascinating spin dynamics, spin-textures, and density patterns \cite{PhysRevA.108.023308}. They have also led to the exploration of various intriguing phenomena such as spin Faraday patterns \cite{PhysRevLett.128.210401}, dynamical stabilization \cite{hoa13}, spin-squeezing \cite{sai08, hoa16}, Shapiro resonances \cite{evr19, ima21}, parametric resonances \cite{pen21}, chaotic dynamics \cite{PhysRevA.81.023619,PhysRevResearch.6.L032030,LIU2022106091}, turbulence \cite{PhysRevA.108.013318,PhysRevA.108.043309}, the quantum walk in momentum space \cite{dad18, dad19} and dynamical phase transitions \cite{PhysRevA.109.043309}. Typically, in experiments involving scalar BECs, the trap frequencies \cite{PhysRevLett.98.095301,PhysRevLett.128.210401} or the interaction parameters, such as the $s$-wave scattering length \cite{PhysRevX.9.011052, zhang20, PhysRevLett.127.113001,PhysRevX.15.011026,PhysRevLett.121.243001}, are periodically modulated. In spinor condensates, the magnetic fields, which are usually used to manipulate internal spin states through linear and quadratic Zeeman effects can also be modulated \cite{PhysRevResearch.6.L032030,PhysRevResearch.1.033132,PhysRevA.108.013318}. Moreover, spinor condensates are characterized by multiple $s$-wave scattering lengths \cite{sad06, sta13, kaw12}, which can be varied independently of each other \cite{zha09}. For instance, a spin-1 Bose gas features two $s$-wave scattering lengths, $a_0$ and $a_2$, offering multiple modulation scenarios by driving either $a_0$,  $a_2$ or both simultaneously. Additionally, the existence of different magnetic ground state phases \cite{kaw12} offers more options for selecting the initial state for the driven dynamics.

The driven dynamics of a spinor condensate critically depends on the initial magnetic state and which scattering length is being modulated \cite{PhysRevA.108.023308}. The ferromagnetic phase is found to be immune to the modulation of $a_0$, whereas $a_2$ modulation results in dynamics similar to that of a driven scalar condensate \cite{sta02, PhysRevLett.98.095301}. In contrast, an initial polar quasi-two-dimensional (Q2D) condensate exhibits a competition between density modulations and spin-mixing dynamics, which characterizes population transfer between magnetic sublevels. This competition can result in the formation of polar-core vortices when spin-mixing dynamics dominates. When both scattering lengths are modulated, the polar phase can display competing instabilities, leading to superpositions of density patterns or correlation functions featuring two distinct wavelengths. Generally, we see that the features and wavelengths involved in the driven dynamics can be predicted from the Bogoliubov spectrum of the initial state, which are analogous to that of scalar condensates \cite{sta02, PhysRevLett.98.095301,PhysRevA.81.033626,PhysRevA.76.063609,PhysRevA.105.063319,PhysRevA.95.043618,PhysRevA.86.023620,PhysRevA.109.033309,PhysRevLett.128.210401,PhysRevA.110.053318,Wan_2024}.

In this paper, we analyze the spin dynamics and the formation of transient Faraday patterns |both spin and density| in quasi-one-dimensional (Q1D) \cite{Q1Dspinorexpt} and Q2D \cite{PhysRevA.101.023613} spin-1 condensates initialized in an anti-ferromagnetic (AFM) phase \cite{PhysRevA.86.061601}, under the parametric driving of the scattering lengths, $a_0$ and $a_2$. The condensate is also subjected to a quadratic Zeeman field (QZF). We consider an AFM phase with equal population in $m=+1$ and $m=-1$ components and its Bogoliubov spectrum features three excitation branches. Among these, two branches are gapless, while the third branch is gapped for a finite QZF. One of the gapless modes corresponds to density excitations, whereas the other is associated with spin waves. The gapped spin mode is related to the transfer of population from $m=\pm 1$ states to the $m=0$ state, and in particular, pairs of bosons with opposite momenta are generated in $m=0$. First, we analyze the effects of $a_0$ modulation. When half of the modulation frequency of scattering lengths lies below the gapped mode, only the two gapless modes are excited, leading to the transient density and spin Faraday patterns. In Q1D, the resulting spin texture is characterized by periodic domains of opposite $z$-polarizations. In contrast, when driven above the gap, the population transfer to the $m=0$ component also occurs, leading to a random spin texture in which the spin vectors are randomly oriented along the condensate axis. When extending to the Q2D spin-1 condensate, distinct features emerge compared to their Q1D counterparts. For instance, in the below-gap case, the spin texture is characterized by ferromagnetic patches with opposite polarizations, irregular shapes, and sizes. When half modulation frequency lies above the gap, we observe the formation of anomalous vortices and antivortices, which do not exhibit phase winding in individual magnetic components unlike polar-core vortices \cite{PhysRevA.110.L061303}. One striking observation is that in Q1D, the spatial spin-spin correlations exhibit a Gaussian envelope, which is unexpected as 1D spin systems typically exhibit either an exponential or a power law behavior. In contrast, in Q2D, the correlations display a Bessel function behavior. Further, we analyze the intriguing scenario of competing instability, where two momenta of the same Bogoliubov mode compete. This phenomenon arises when both scattering lengths are modulated simultaneously at different frequencies and with appropriately chosen amplitudes. Strikingly, various scenarios of competing instability can be achieved by simply varying the QZF while keeping all other parameters constant. This leads to a complex behavior in the population transfer from $m=\pm 1$ components to the $m=0$ component as the QZF is varied, especially exhibiting abrupt jumps in the total population of the $m=0$ component obtained at fixed times.

The paper is structured as follows. In Sec.~\ref{model}, we discuss the setup, corresponding spinor Gross-Pitaevskii equations and Bogoliubov excitations of homogeneous spin-1 condensate in the AFM phase. The Mathieu equations governing the stability of driven spin-1 condensate are discussed in Sec.~\ref{tm}. The dynamics under $a_0$ modulation is discussed in Sec.~\ref{a0m} and in particular, that of Q1D and Q2D condensates are analyzed in Secs.~\ref{q1d} and \ref{q2d}, respectively. Similarly, the dynamics under $a_2$ modulation is discussed in Sec.~\ref{a2m}, and the dynamics under simultaneous modulation of $a_0$ and $a_2$ is discussed in Sec.~\ref{a02m}. Finally, we summarize and provide an outlook in Sec.~\ref{sok}.

\section{setup, model and Bogoliubov excitations}
\label{model}

%
\begin{figure}
\centering
\includegraphics[width= 1.\columnwidth]{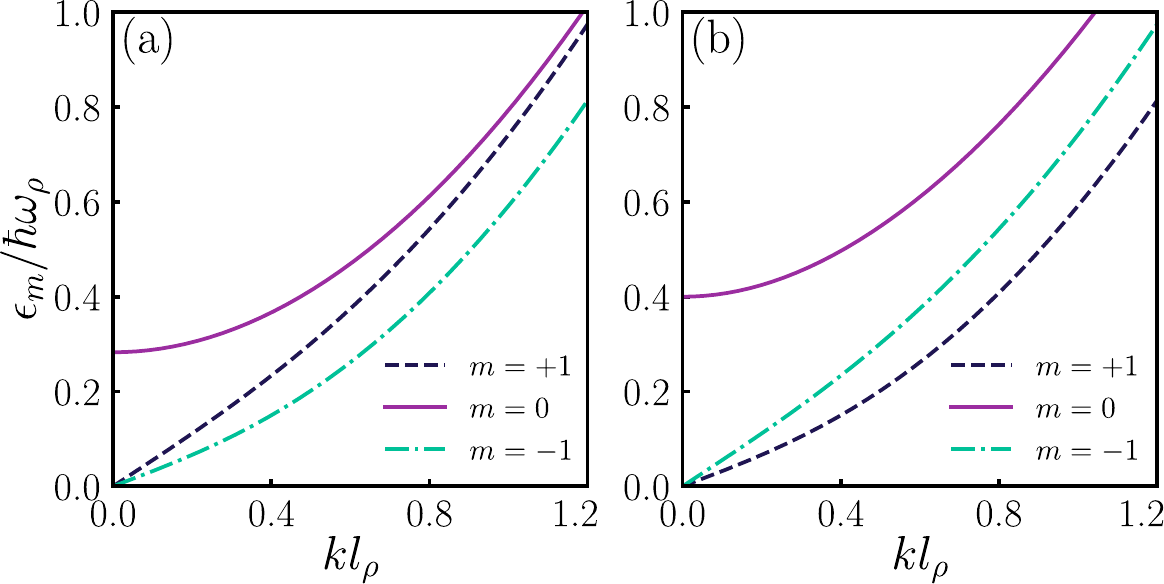}
\caption{\small{(color online). Bogoliubov spectrum of a quasi-1D spin-1 BEC with the spin order parameter $(1, 0, 1)/\sqrt{2}$ for (a) $q/\hbar\omega_\rho=-0.2$, $c_0n_H^{1D}/2\pi l_\rho^2\hbar\omega_\rho=0.4$, $c_1n_H^{1D}//2\pi l_\rho^2\hbar\omega_\rho=0.2$ and (b) $q/\hbar\omega_\rho=-0.2$, $c_0n_H^{1D}/2\pi l_\rho^2\hbar\omega_\rho=0.2$, $c_1n_H^{1D}/2\pi l_\rho^2\hbar\omega_\rho=0.4$, where $l_\rho=\sqrt{\hbar/M\omega_\rho}$. $k$ is the quasi-particle momentum along the axial $z$ direction.}}
\label{fig:1} 
\end{figure}

We consider a spin-1 BEC with each boson having a mass $M$ governed by the coupled Gross-Pitaevskii equations (GPEs),
\begin{eqnarray}
\label{m1}
i\hbar\frac{\partial}{\partial t}\psi_1 &=& \left[\hat{\mathcal{H}_0 }+q+c_1 F_z(\bm r)\right]\psi_1 (\bm r)+\frac{ c_1}{\sqrt 2}F_-(\bm r)\psi_0(\bm r),\label{GPE3D-a}\nonumber\\\\
\label{m2}
i\hbar\frac{\partial}{\partial t}\psi_0 &=& \hat{\mathcal{H} }_0\psi_0(\bm r) +\frac{c_1}{\sqrt 2}\left[F_+(\bm r)\psi_1(\bm r) +F_-(\bm r)\psi_{-1}(\bm r)\right],\label{GPE3D-b}\nonumber\\ \\
\label{m3}
i\hbar\frac{\partial}{\partial t}\psi_{-1} &=& \left[\hat{\mathcal{H}_0}+q-c_1 F_z(\bm r)\right]\psi_{-1}(\bm r) +\frac{c_1}{\sqrt 2}F_+(\bm r)\psi_0(\bm r) \label{GPE3D-c},
\end{eqnarray}
where $\psi_m(\bm r)$ is the wave function of the $m$th Zeeman-component,  $ \hat{\mathcal{H}_0}= -(\hbar^2/2M)\nabla^2 +V_{ext}(r)+c_0 n(\bm r) $ with $c_0$ being the spin-independent interaction strength, and $n(\bm r)=\sum_mn_m(\bm r)$ is the total density, where $n_m(\bm r)=|\psi_m({\bm r})|^2$. The components of spin density are given by $F_\nu(\bm r)=\sum_{m,m'}\psi^*_m(\bm r)(f_\nu)_{mm'}\psi_{m'}(\bm r)$ with $f_\nu$ being the $\nu$th component of the spin-1 matrices and $ F_\pm=F_x(\bm r)\pm i F_y(\bm r)$. Writing explicitly, we have $F_z(\bm r)=n_1(\bm r)-n_{-1}(\bm r)$, $F_x(\bm r)=[\psi_1^*\psi_0+\psi_0^*(\psi_1+\psi_{-1})+\psi_{-1}^*\psi_0]/\sqrt{2}$ and $F_y(\bm r)=i[-\psi_1^*\psi_0+\psi_0^*(\psi_1-\psi_{-1})+\psi_{-1}^*\psi_0]/\sqrt{2}$. The parameter $q$ provides the strength of  the QZF, while $c_1$ is the spin-dependent interaction strength. The external trapping potential, $V_{ext}(r)=M\omega_z^2z^2/2$ and  $V_{ext}(r)=M\omega_\rho^2(x^2+y^2)/2$, respectively for the Q2D and Q1D condensates. We assume that the trap frequencies are significantly large compared to the Q1D and Q2D chemical potentials, allowing us to treat the transverse wave functions as the Gaussian ground states of the harmonic trap. By integrating over the strongly confined directions, we arrive at the low-dimensional coupled GPEs, which retain the same form as Eqs.~(\ref{m1})-(\ref{m3}), but with rescaled interaction parameters. These equations are then solved numerically to investigate the modulation-induced dynamics, starting from a homogeneous density for the $m=\pm 1$ components, embedded with a TWA noise in all three components \cite{sai08,PhysRevA.108.023308}. We set $c_0$ and $c_1$ to be greater than zero and $q<0$ such that the initial AFM phase becomes the ground state of the spin-1 condensate  \cite{kaw12}.


The dispersion relations of the Bogoliubov modes of a three-dimensional (3D) homogeneous spin-1 condensate with density $n_H$ and a spin order parameter $(1, 0, 1)/\sqrt{2}$ (AFM phase), having the chemical potential, $\mu=c_0n_H+q$ are given by \cite{kaw12}, 
\begin{eqnarray}
\label{b1}
\epsilon_{0}(k)=\sqrt{(E_k-q)^2+2 c_1n_H(E_k-q)}\\
\label{b2}
\epsilon_{+1}(k)=\sqrt{E_k(E_k+2c_0n_H)} \\
\label{b3}
\epsilon_{-1}(k)=\sqrt{E_k(E_k+2c_1n_H)},
\end{eqnarray}
where $k$ is the quasi-momentum and $E_k=\hbar^2k^2/2M$. The modes given in Eqs.~(\ref{b1})-(\ref{b3}) take the same mathematical form also for Q1D and Q2D homogeneous condensates. The mode $\epsilon_{0}$ is gapped by the QZF, whereas $\epsilon_{\pm 1}$ are gapless and is independent of $q$. The gap of $\epsilon_{0}$ mode is given by $\hbar\Delta=\sqrt{q^2-2qc_1n_H}$. In Fig.~\ref{fig:1}, we show the Bogoliubov spectrum of a Q1D homogeneous BEC with density $n_H^{1D}=2\pi l_\rho^2n_H$, where $l_\rho=\sqrt{\hbar/m\omega_\rho}$. When $c_0>c_1$, the frequency of the density mode, $\epsilon_{+1}$, is greater than that of the magnon mode  $\epsilon_{-1}$ [see Fig.~\ref{fig:1}(a)], and vice versa [see Fig.~\ref{fig:1}(b)]. The density mode represents the excitations in the total density $n(z, t)=\sum_mn_m(z, t)$ and the magnon mode corresponds to the spatial oscillations of the local magnetization density, $F_z(z, t)$. The gapped $\epsilon_{0}$ mode characterizes the population transfer between $m=\pm 1$ and $m=0$ magnetic components, which we refer to as spin dynamics. Also, note that this is in stark contrast with the Bogoliubov spectrum of the polar phase, which contains one gapless density mode and two gapped modes, and hence, we anticipate qualitatively different driven dynamics from that of the driven polar phase \cite{PhysRevA.108.023308}. In the numerical calculations of GPEs, the initial AFM phase is embedded with a noise field, for instance, in 3D it reads as,
	\begin{eqnarray}
		\bm \delta ({\bm r}) &=&  \frac{1}{\sqrt{V}}\sum_{{\bm k}}\left[
		\begin{pmatrix}
			\left(\beta_{\bm k,1} \bar u_{\bm k,1} +\beta_{\bm k,-1} \bar u_{\bm k,-1}\right)/\sqrt{2}  \\ \beta_{\bm k,0} \bar u_{\bm k,0} \\ \left(\beta_{\bm k,1} \bar u_{\bm k,1} -\beta_{\bm k,-1} \bar u_{\bm k,-1}\right) /\sqrt{2} 
		\end{pmatrix} e^{-i\bm k\cdot \bm r} \right. \nonumber \\ 
		&&\left. + 
		\begin{pmatrix}
			\left(\beta^*_{\bm k,1} \bar v_{\bm k,1} +\beta^*_{\bm k,-1} \bar v_{\bm k,-1}\right) /\sqrt{2}  \\ \beta^*_{\bm k,0} \bar v_{\bm k,0} \\ \left(\beta^*_{\bm k,1} \bar v_{\bm k,1} -\beta^*_{\bm k,-1} \bar v_{\bm k,-1} \right)/\sqrt{2} 
		\end{pmatrix}e^{i\bm k\cdot \bm r}\right]
	\end{eqnarray}
where $k=|\bm k|$, $\bar u_{\bm k,1} = \sqrt{[E_k+\bar c_0n_H+\epsilon_{1}(k)]/2\epsilon_{1}(k)}$, $\bar u_{\bm k,0} = \sqrt{[E_k-q+\bar c_1n_H+\epsilon_{0}(k)]/2\epsilon_{0}(k)}$, $\bar u_{\bm k,-1} = \sqrt{[E_k+\bar c_1n_H+\epsilon_{-1}(k)]/2\epsilon_{-1}(k)}$, and $ \bar v_{\bm k,m} = \sqrt{\bar u_{\bm k,m}^2-1} $. $ \beta_{\bm k,m}$ are random numbers with zero mean and a variance of $1/2$. The 1D and 2D versions of the TWA noise can be written accordingly.


\section{Time modulation of $s$-wave scattering lengths: Mathieu equations}
\label{tm}

\begin{figure}
\centering
\includegraphics[width= 1.\columnwidth]{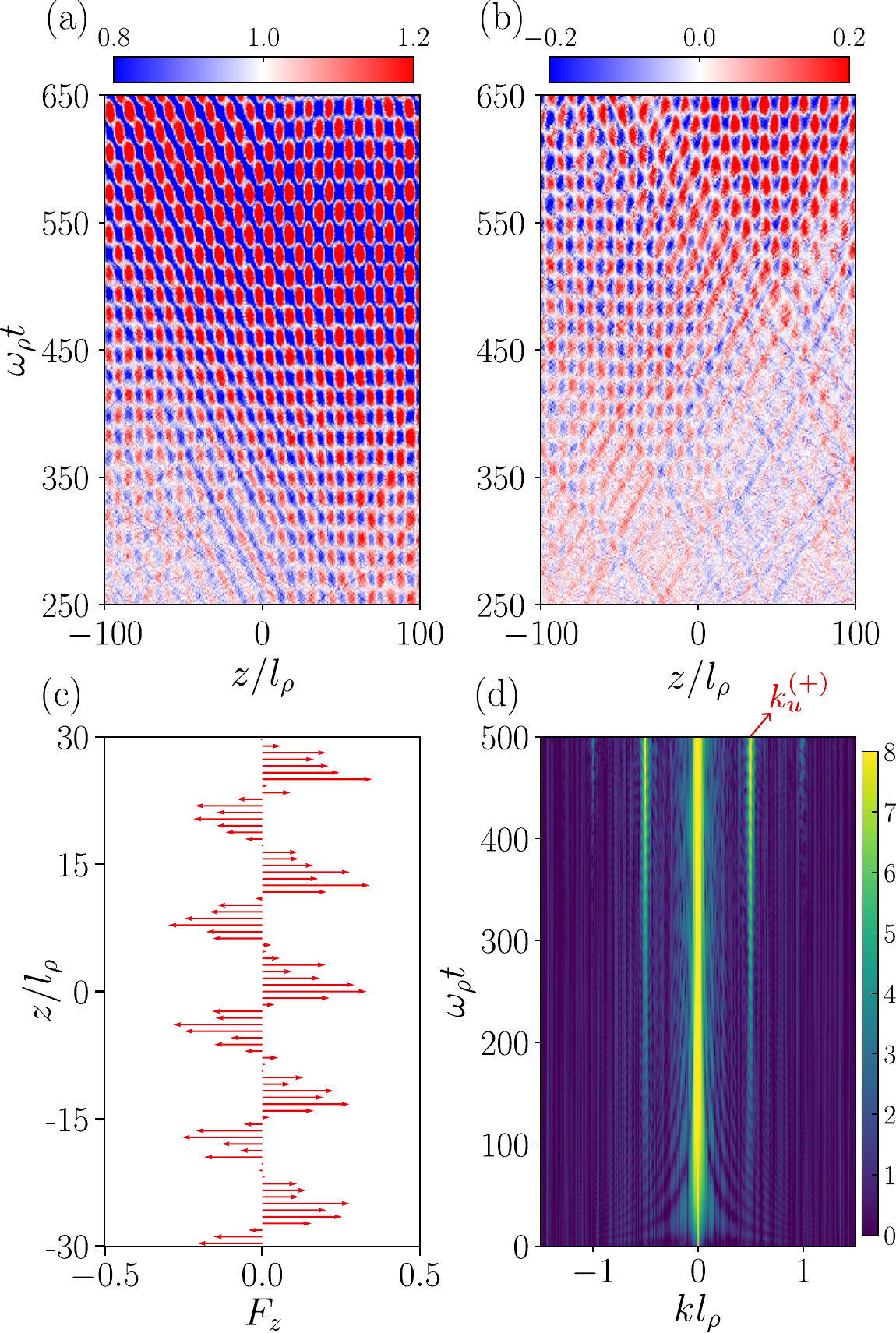}
\caption{\small{(color online). Results for $\omega_0<\Delta$. The (a) density ($n$) and (b) spin ($F_z$) Faraday patterns in a Q1D spin-1 BEC starting from an AFM phase for $\omega_0=0.2\omega_\rho$, $q/\hbar\omega_\rho=-0.2$, $\bar c_{0}n_H^{1D}/2\pi l_\rho^2\hbar\omega_\rho=\bar c_{1}n_H^{1D}/2\pi l_\rho^2\hbar\omega_\rho=0.1$, and $\alpha_0=0.45$. The value of gap is $\Delta=0.283\omega_\rho$. (c) The spin texture at $\omega_\rho t=600$, showing periodic domains of spins pointing up and down along the $z$-axis. (d) The dynamics of condensate density in the momentum space, $\tilde n(k)$ reveals the unstable momentum $k_u^{(+)}$ and the peak at $k=0$ corresponds to the homogeneous density.}}
\label{fig:2} 
\end{figure}

The interaction parameters can be expressed in terms of the $s$-wave scattering lengths as, $c_0=(g_0+2g_2)/3$ and $c_2=(g_2-g_0)/3$ where $g_{\mathcal F}=4\pi\hbar^2a_{\mathcal F}/m$ is related to the scattering length $a_{\mathcal F}$ of the total spin-$\mathcal F$ channel with $\mathcal F\in\{0, 2\}$.  At this point, we consider time dependent scattering lengths, $a_{\mathcal F}(t)=\bar a_{\mathcal F}[1+2\alpha_{\mathcal F}\cos (2\omega_{\mathcal F} t)]$, where $\bar a_{\mathcal F}$ is the mean scattering length, $\alpha_{\mathcal F}$ is the modulation amplitude and $2\omega_{{\mathcal F}}$ is the driving frequency. In the general case where both scattering lengths are modulated, the time-dependent interaction coefficients become
\begin{eqnarray}
 c_0(t)=\bar c_0+(2\alpha_{0}\bar g_0/3)\cos(2\omega_0 t)+(4\alpha_{2}\bar g_2/3)\cos(2\omega_2 t), \\
 c_1(t)=\bar c_1-(2\alpha_0\bar g_0/3)\cos(2\omega_0 t)+(2\alpha_2\bar g_2/3)\cos(2\omega_2 t),
\end{eqnarray}
where $\bar g_{\mathcal F}=4\pi\hbar^2\bar a_{\mathcal F}/m$, $\bar c_0=(\bar g_0+2\bar g_2)/3$ and $\bar c_2=(\bar g_2-\bar g_0)/3$. The homogeneous AFM solution in the presence of modulation is $\bm \psi(t)=\sqrt{n_H}{\bm \zeta}\exp(-i\theta(t)/\hbar)$ where $\bm \psi(t)=(\psi_1, \psi_0, \psi_{-1})^T$, ${\bm \zeta}=(1, 0, 1)^T/\sqrt{2}$ and $\theta(t) = qt + n_H\int_0^t dtc_0(t)$. Introducing
\begin{equation}
\bm \psi(\bm r, t)=\left[\sqrt{n_H}\bm \zeta+\bm w(t)\cos({\bm k}\cdot{\bm r})\right]e^{-i\theta(t)/\hbar},
\end{equation}
in Eqs.~(\ref{m1})-(\ref{m3}) and linearize in $\bm w(t )$, where $\bm w(t)=(w_1, w_0, w_{-1})^T$ is the amplitude of modulations and $\bm w(t)=\bm u(t)+i\bm v(t)$ with $\bm u=(u_1, u_0, u_{-1})^T$ and $\bm v=(v_1, v_0, v_{-1})^T$ being real-valued vectors, we arrive at the following Mathieu-like second-order differential equations: 
\begin{eqnarray}
	\label{me1}
		\frac{d^2 u_{+1}}{dt^2}+\frac{1}{\hbar^2}\left[E_k(E_k+2c_0(t)n_H )\right]u_{+1} &=& 0,\label{mat1} \\	
			\label{me2}
		\frac{d^2 u_{0}}{dt^2}+\frac{1}{\hbar^2}\left[(E_k-q)(E_k-q+2c_1(t)n_H )\right]u_{0} &=& 0,\label{mat2} \\
					\label{me3}
		\frac{d^2 u_{-1}}{dt^2}+\frac{1}{\hbar^2}\left[E_k(E_k+2c_1(t)n_H )\right]u_{-1} &=& 0 \label{mat3}.
	\end{eqnarray}
Depending on the modulation scheme, we can insert appropriate $c_0(t)$ and $c_1(t)$ in Eqs.~(\ref{me1})-(\ref{me3}) and analyze the stability properties, particularly the wave number selection of the Faraday patterns in the driven condensate. Numerically, we identify the unstable momenta using the condensate density in the momentum space, $\tilde n(k)=\sum_m|\tilde \psi_m(k)|^2$, where $\tilde \psi_m(k)$ is the Fourier transform of the condensate wave function, and compare with those obtained from the stability analysis of Mathieu equations. 

\section{Dynamics under $a_0$ modulation}
\label{a0m}

Under the $a_0$ modulation ($\alpha_0\neq 0$ and $\alpha_2=0$), Eqs.~(\ref{me1})-(\ref{me3}) become
\begin{eqnarray}
	\label{mea01}
		\frac{d^2 u_{+1}}{dt^2}+\frac{1}{\hbar^2}\left[\epsilon_{+1}(k)^2+\frac{4\alpha_0\bar g_0n_HE_k}{3}\cos(2\omega_0t)\right]u_{+1} = 0, \\
			\label{mea02}
		\frac{d^2 u_{0}}{dt^2}+\frac{1}{\hbar^2}\left[\epsilon_{0}(k)^2-\frac{4\alpha_0\bar g_0n_H(E_k-q)}{3}\cos(2\omega_0t)\right]u_{0}= 0,  \\
			\label{mea03}	
		\frac{d^2 u_{-1}}{dt^2}+\frac{1}{\hbar^2}\left[\epsilon_{-1}(k)^2-\frac{4\alpha_0\bar g_0n_HE_k}{3}\cos(2\omega_0t)\right]u_{-1} =0. \ \ 	
	\end{eqnarray}
For a fixed driving frequency $\omega_0$ and in the limit $\alpha_0\to 0$, the resonances $\epsilon_{k, m=0,  \pm 1} = \hbar \omega_0$ provide three unstable momenta $k_u^{(0)}$ and $k_u^{(\pm)}$ with Floquet exponents $\sigma^{(0)}\simeq n_H(E_{k_u^{(0)}}-q)\alpha_0\bar g_0/3\hbar^2\omega_0$ and $\sigma^{(\pm)}\simeq  n_HE_{k_u^{(\pm)}}\alpha_0\bar g_0/3\hbar^2\omega_0$ \cite{PhysRevA.81.033626}, respectively. The larger the magnitude of the Floquet exponent, the more unstable the corresponding mode is. Below the gap, only the two momenta, $k_u^{(+)}$ and $k_u^{(-)}$ are relevant. The momentum dependence of Floquet exponents ($\propto k^2$) indicates that, for a fixed driving frequency, the largest among $k_u^{(+)}$ and $k_u^{(-)}$ becomes more unstable, which is provided by the lowest lying mode in the Bogoliubov spectrum. The latter is the magnon mode for $c_0>c_1$ [see Fig.~\ref{fig:1}(a)] and density mode for $c_0<c_1$ [see Fig.~\ref{fig:1}(b)]. Since the most unstable momentum determines the characteristics of the post-instability dynamics, we anticipate contrasting dynamics for $c_0\gg c_1$ and $c_0\ll c_1$. Obtaining $k_u^{(0)}$ and $k_u^{(\pm)}$ from resonance conditions, we rewrite the Floquet exponents as
 \begin{eqnarray}
 \label{s0}
 \sigma^{(-)}=\sigma^{(0)}=\frac{\alpha_0n_H\bar g_0}{3\hbar^2\omega_0}\left[\sqrt{(n_H\bar c_1)^2+\hbar^2\omega_0^2}-n_H\bar c_1\right],\\
 \sigma^{(+)}=\frac{\alpha_0 n_H\bar g_0}{3\hbar^2\omega_0}\left[\sqrt{(n_H\bar c_0)^2+\hbar^2\omega_0^2}-n_H\bar c_0\right],
 \label{s1}
 \end{eqnarray}
 interestingly, which are independent of $q$, the QZF. When $\bar c_0=\bar c_1$, we have $\sigma^{(0)}=\sigma^{(\pm)}$, indicating that all three modes may play an equal role in the dynamics. If that is the case, spin dynamics (population transfer between sublevels), transient density, and spin Faraday patterns would emerge simultaneously. Note that the instability analysis based on Eqs.~(\ref{mea01})-(\ref{mea03}) is independent of the dimension of the condensate, whereas the post-instability dynamics can depend critically on it. At longer times, the condensates are destroyed via heating. First, we will discuss the dynamics of Q1D and Q2D condensates under $a_0$ modulation.
\subsection{Q1D condensate}
\label{q1d}

\begin{figure}
\centering
\includegraphics[width= 1.\columnwidth]{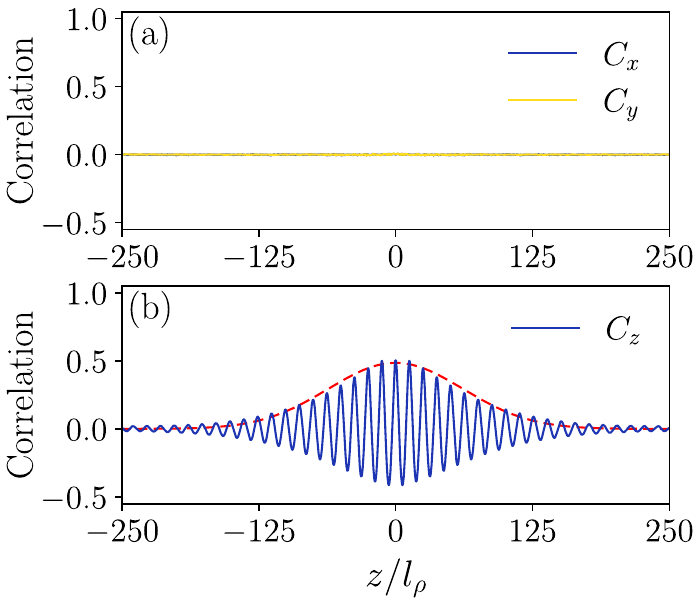}
\caption{\small{(color online). Q1D spin-spin correlations of for $\omega_0<\Delta$: (a) $C_x$ and $C_y$, and (b) $C_z$, for the same parameters as in Fig.~\ref{fig:2}. $C_z$ exhibits a Gaussian envelope. The time average is taken from $t_i=300\omega_\rho^{-1}$ to $t_f=650\omega_\rho^{-1}$. We further take an average over 40 realizations of TWA noises. The dashed line shows the Gaussian fit for the envelope. The central peak, $C_\alpha(z=0)=1$ is removed for clarity.}}
\label{fig:3} 
\end{figure}
In the following, we consider two cases of driving frequencies, one below the gap ($\omega_0<\Delta$) and the other above the gap ($\omega_0>\Delta$), and analyze the driven dynamics for $\bar c_0= \bar c_1$, $\bar c_0\ll \bar c_1$, and $\bar c_0\gg \bar c_1$.  

\subsubsection{$\omega_0<\Delta$}
\label{btg}
Below the gap ($\omega_0<\Delta$), the relevant modes are $\epsilon_+(k)$ and $\epsilon_-(k)$. They yield two unstable momenta, $k_u^{(+)}$ and $k_u^{(-)}$ corresponding to density and magnon excitations. In Fig.~\ref{fig:2}, we show the results for $\bar c_0= \bar c_1$, where we observed the simultaneous formation of density [$n(z,t)$, see Fig.~\ref{fig:2}(a)] and axial spin [$F_z(z,t)$, see Fig.~\ref{fig:2}(b)] Faraday patterns. Since $\epsilon_+(k)=\epsilon_-(k)$ for $\bar c_0= \bar c_1$, we have $k_u^{(+)}=k_u^{(-)}$. Hence, the density and spin patterns exhibit the same wavelength or periodicity, provided by $1/k_u^{(+)}$. The unstable momenta $\pm k_u^{(+)}$ can be identified from the condensate density in the momentum space [see Fig.~{\ref{fig:2}(c)}]. The peak at $k=0$ corresponds to the homogeneous condensate density. Since the mode $\epsilon_0(k)$ is not excited in this case, there is no population transfer to the $m=0$ state from the $m=\pm 1$ states, leaving the transverse magnetizations, $F_{x, y}(z, t)\sim 0$. Hence, we have a spin texture shown in Fig.~\ref{fig:2}(c), which characterizes periodic domains of opposite $z$-polarizations. Also, it can be equivalently identified as a dynamical array of domain walls or kinks. At the kinks, the spin density vanishes. 

Further, we look at the scaled spin-spin correlations,
\begin{equation}
C_{\alpha}(z)=\frac{1}{N_{\alpha}\tau}\int_{t_i}^{t_f} dt \int dz'F_{\alpha}(z+z', t)F_{\alpha}(z', t),
\label{cor1}
\end{equation}
where $\alpha\in\{x, y, z\}$ and $N_{\alpha}=\int dzF_{\alpha}(z)^2$. The initial instant $t_i$ is the time at which the pattern begins to appear, $t_f$ is selected just before the condensate is destroyed by heating, and $\tau=t_f-t_i$. $C_{\alpha}(z)$ for $\omega_0<\Delta$ is shown in Fig.~\ref{fig:3}, which reveals no spin correlations along the $x$ and $y$ directions as expected [see Fig.~\ref{fig:3}(a)], whereas, along the $z$ axis, we observe that the correlations exhibit a Gaussian envelope [see Fig.~\ref{fig:3}(b)]. To the best of our knowledge, no spin system exhibits a Gaussian profile for the spatial spin-spin correlations, whereas the temporal (autocorrelations) profile exhibiting Gaussian behavior at infinite temperature is well known for an XX spin chain \cite{PhysRevB.52.4319}. Generally, the envelope of spatial correlations is characterized by either exponential decay or power-law behavior. The oscillations in $C_{z}(z)$ [see Fig.~\ref{fig:3}(b)] is characterized by the unstable momentum $k_u^{(+)}$.  

When $\bar c_0 \neq \bar c_1$, the degeneracy between $\epsilon_+(k)$ and $\epsilon_-(k)$ is broken [see Fig.~{\ref{fig:1}}], which allows for the selective excitation of one of the two modes through periodic driving. For instance, when $\bar c_0 \ll \bar c_1$, we have $\sigma^{(+)}\gg \sigma^{(-)}$, indicating that the transient density pattern occurs while the spin pattern is suppressed. In other words, the density mode becomes softer and the magnon mode stiffens. In contrast, when $\bar c_0 \gg \bar c_1$, the spin pattern becomes prominent while the density pattern is suppressed. Numerical results have further confirmed these predictions.

\subsubsection{$\omega_0>\Delta$}
\label{atg}

\begin{figure}
\centering
\includegraphics[width= 1.\columnwidth]{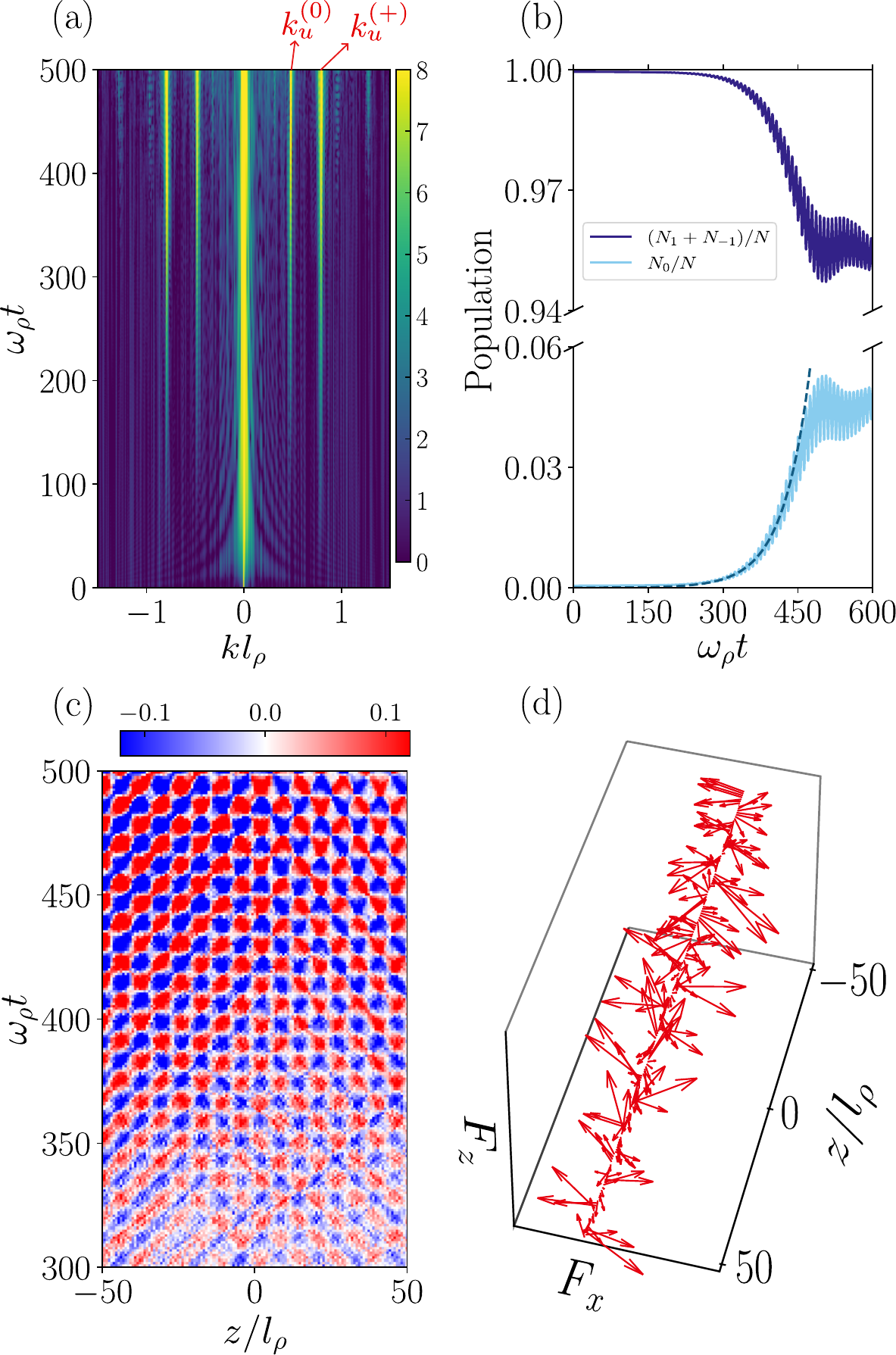}
\caption{\small{(color online). Results for $\omega_0>\Delta$. (a) The dynamics of condensate density in the momentum space, $\tilde n(k)$ reveals the unstable momentum $k_u^{(+)}$ and $k_u^{(0)}$. The peak at $k=0$ corresponds to the homogeneous density. (b) The spin dynamics exhibiting population transfer from $m=\pm 1$ components to $m=0$ component due to the unstable $\epsilon_{0}(k)$ mode, where $N_m(t)=\int dz|\psi_m(z, t)|^2dz$ and $N=\sum_mN_m$. The dashed line is $\exp[\sigma^{(0)}t]$, which captures the initial exponential population growth in $m=0$ accurately and also indicates the linear regime of the dynamics. (c) shows the dynamics of $F_x$ and (d) the snap shot of random spin texture at $\omega_\rho t=500$. The values of system parameters are $\omega_0=0.4\omega_\rho$, $q/\hbar\omega_\rho=-0.2$, $\bar c_{0}n_H^{1D}/2\pi l_\rho^2\hbar\omega_\rho=\bar c_{1}n_H^{1D}/2\pi l_\rho^2\hbar\omega_\rho=0.1$, and $\alpha_0=0.45$. The value of gap is $\Delta=0.283\omega_\rho$.}}
\label{fig:4} 
\end{figure}

\begin{figure}
\centering
\includegraphics[width= .94\columnwidth]{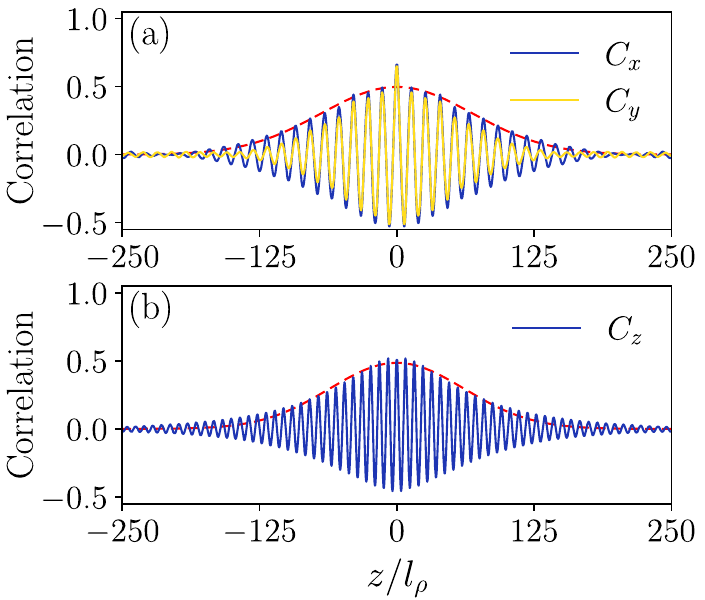}
\caption{\small{(color online). Spin-spin correlations for $\omega_0>\Delta$: (a) $C_x$ and $C_y$, and (b) $C_z$, for the same parameters as in Fig.~\ref{fig:4}. All three correlations exhibit a Gaussian envelope. The time average is taken from $t_i=200\omega_\rho^{-1}$ to $t_f=600\omega_\rho^{-1}$. We further take an average over 40 realizations of TWA noises. The dashed line shows the Gaussian fit for the envelope and the central peak, $C_\alpha(z=0)=1$ is removed for clarity.}}
\label{fig:5} 
\end{figure}

\begin{figure}
\centering
\includegraphics[width= 1. \columnwidth]{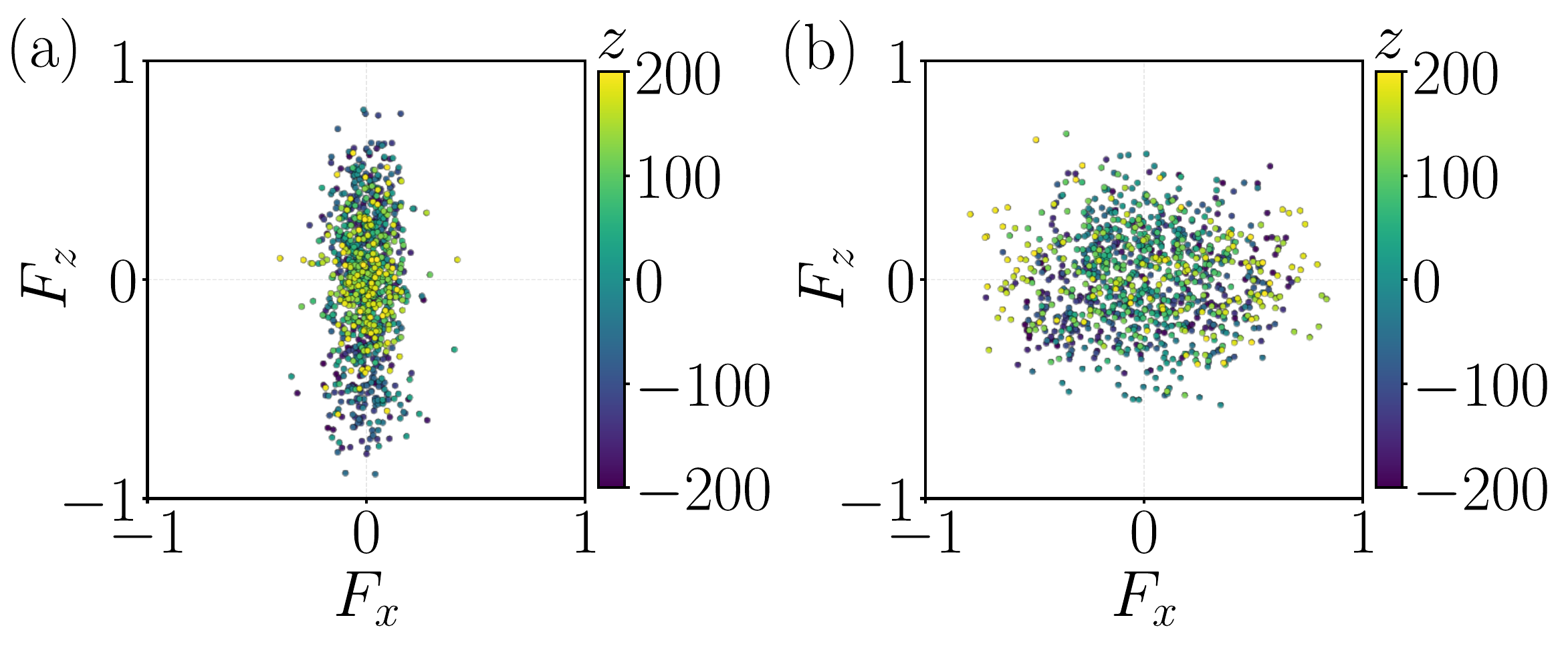}
\caption{\small{(color online). Projection of spin vector on to the $F_x-F_z$ plane for (a) $\omega_0<\Delta$ and (b) $\omega_0>\Delta$ obtained respectively at $\omega_\rho t=600$ and $\omega_\rho t=500$. The parameters of (a) are the same as that of Fig.~\ref{fig:2} and that of (b) are the same as that of Fig.~\ref{fig:4}.}}
\label{fig:6} 
\end{figure}

When half of the driving frequency $\omega_0$ exceeds the gap $\Delta$, all three modes become dynamically unstable, giving us three unstable momenta, $k_u^{(\pm)}$ and $k_u^{(0)}$. For $\bar c_0= \bar c_1$ (density and magnon modes are degenerate),   $k_u^{(+)}=k_u^{(-)}$, and both $k_u^{(+)}$ and $k_u^{(0)}$ populations grow at the same rate, as seen in the momentum space of the condensate density [see Fig.~\ref{fig:4}(a)]. In contrast to the case of $\omega_0<\Delta$, we observe a transfer of population to the $m=0$ state due to the unstable $\epsilon_{0}(k)$ mode, as shown in Fig.~\ref{fig:4}(b). The latter makes $F_{x,y}\neq 0$ and the dynamics of $F_x$ is shown in Fig.~\ref{fig:4}(c), which reveals a pattern identical to the spin Faraday pattern shown previously for $\omega_0<\Delta$. Note that the oscillations in $F_z$ are governed by the wave number $k_u^{(-)}$ whereas that of $F_x$ are quantified by $k_u^{(0)}$. The growth of fluctuations in $F_x$ and $F_y$ also critically depends on the initial phase difference between the components $m=1$ and $m=-1$. Since we took that to be zero, initially, $F_x$ grows faster than $F_y$. As expected from the linear stability analysis, the initial exponential growth of population $N_0$ in $m=0$ state is entirely determined by the Mathieu exponent $\sigma^{(0)}$ given in Eq.~(\ref{s0}) [depicted by the dashed line in Fig.~\ref{fig:4}(b)]. In this case, the spin vector reveals a random spin texture, as shown in Fig.~\ref{fig:4}(d), and the spin-spin correlations along all three directions are found to exhibit a Gaussian profile as shown in Fig.~\ref{fig:5}. Note that the oscillations in $C_z$ are governed by $k_u^{(-)}$ whereas that of $C_{x, y}$ is provided by $k_u^{(0)}$ as anticipated. Ultimately, the nature of spin textures can be revealed by projecting the spin vector at each point along the $z$-axis onto the $F_x-F_z$ plane [see Fig.~\ref{fig:6}]. For $\omega_0<\Delta$, this projection generates a stripe, whereas for $\omega_0>\Delta$, it forms a disc shape, indicating the random orientations that vary from position to position.

As in the case of $\omega_0<\Delta$, the degeneracy of density and magnon modes is broken for $\bar c_0 \neq \bar c_1$, and we can again selectively excite density or spin patterns. In contrast with the $\omega_0<\Delta$ case, for $c_1\ll c_0$, the spin Faraday pattern is always accompanied by population transfer from $m=\pm 1$ to $m=0$ state, since $\sigma^{(-)}=\sigma^{(0)}$.

\subsection{Q2D condensate}
\label{q2d}

\begin{figure}
\centering
\includegraphics[width= 1.\columnwidth]{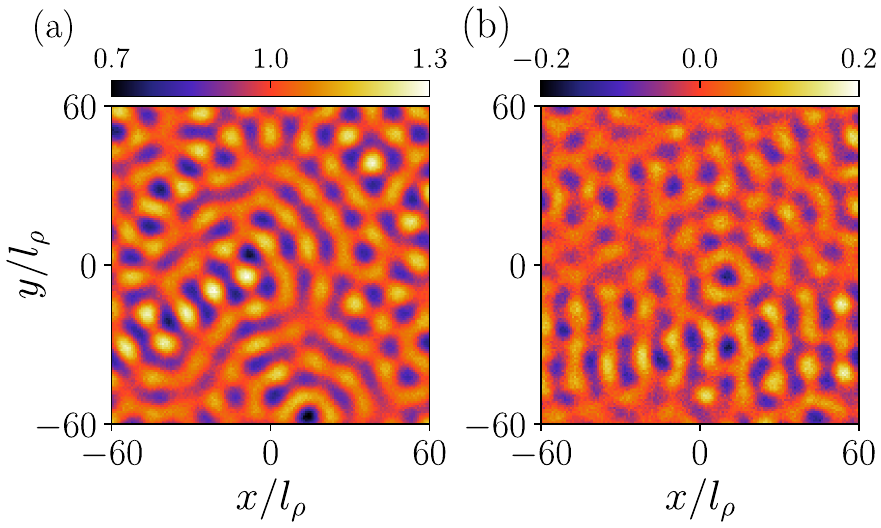}
\caption{\small{(color online). Two dimensional Faraday patterns for $\omega_0<\Delta$. (a) density ($n$) and (b) spin ($F_z$) patterns $\omega_0=0.2\omega_z$, $q/\hbar\omega_z=-0.2$, $\bar c_{0}n_H^{2D}/\sqrt{2\pi} l_z\hbar\omega_z=\bar c_{1}n_H^{2D}/\sqrt{2\pi} l_z\hbar\omega_z=0.1$, and $\alpha_0=0.4$. The value of gap is $\Delta=0.283\omega_z$. The snapshot of the density pattern is taken at $\omega_z t=680$ and that of spin pattern is $\omega_z t=780$.}}
\label{fig:7} 
\end{figure}

\begin{figure}
\centering
\includegraphics[width= 1.\columnwidth]{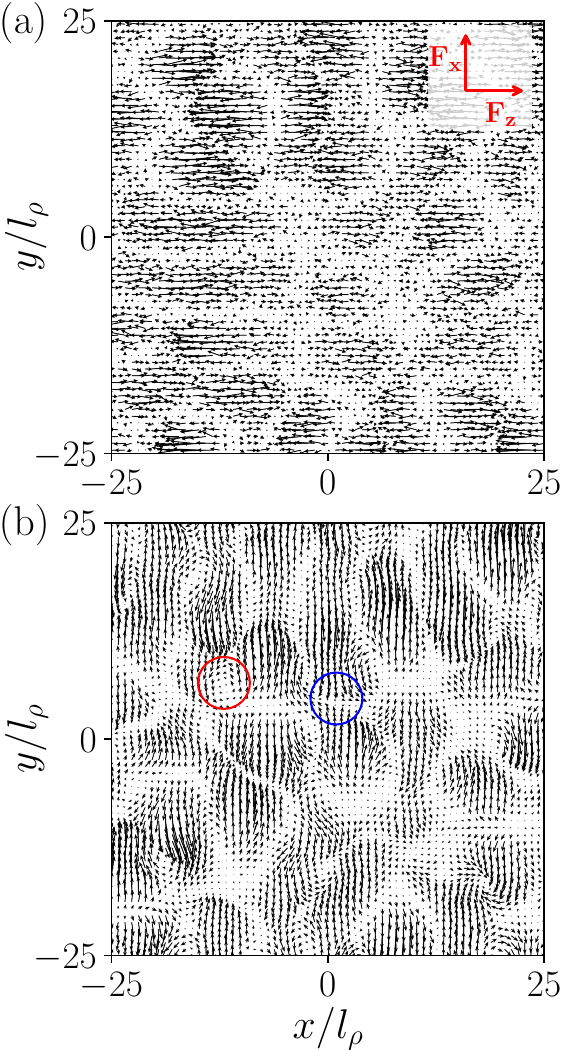}
\caption{\small{(color online). The spin textures in the $F_x$-$F_z$ plane of a driven Q2D spin-1 condensate starting from the AFM phase for (a) $\omega_0<\Delta$ (below gap) and  (b) $\omega_0>\Delta$ (above gap). (a) displays ferromagnetic patches for $\omega_0=0.2\omega_z$, taken at $\omega_z t=780$ (b) Anomalous vortices and antivortices (marked by circles) for $\omega_0=0.4\omega_z$, taken at $\omega_z t=610$. The other parameter values are same as in Fig.~\ref{fig:6}.}}
\label{fig:8} 
\end{figure}

\begin{figure}
\centering
\includegraphics[width= 1.\columnwidth]{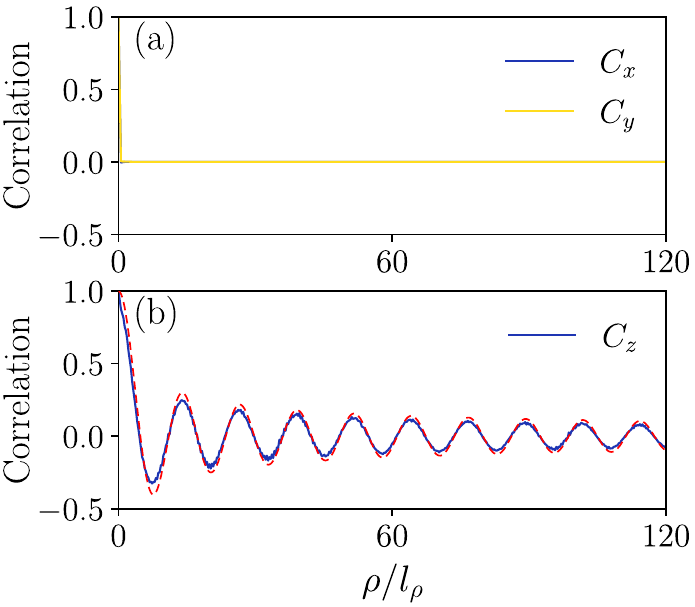}
\caption{\small{(color online). Spin-spin correlations for $\omega_0<\Delta$: (a) $C_x$ and $C_y$, and (b) $C_z$, for the same parameters as in Fig.~\ref{fig:6} at $\omega_z t=830$. The results are obtained from an average over 10 realizations of TWA noises. The dashed line shows the Bessel function fit to the numerical results. }}
\label{fig:9} 
\end{figure}

In the following, we extend the above analysis to a homogeneous quasi-2D condensate with density $n_H^{2D}=\sqrt{2\pi}l_zn_H$, where $l_z=\sqrt{\hbar/m\omega_z}$. In the below-gap case, some qualitative features remain the same as those of the Q1D condensate. The corresponding density and spin Faraday patterns are shown in Fig.~\ref{fig:7}. The related spin texture is shown in Fig.~\ref{fig:8}(a), in which the local spin vectors predominantly point along either the positive or negative z-axis, resulting in ferromagnetic patches \cite{Bert2011} with opposite polarizations. These magnetic patches exhibit variable and irregular shapes and sizes. In the above gap case, the spin dynamics are qualitatively similar to the Q1D case; however, the spin textures exhibit different characteristics [see Fig.~\ref{fig:8}(b)]. Notably, we identify the formation of anomalous vortices and antivortices \cite{PhysRevA.110.L061303}. They are anomalous in the sense that they do not exhibit phase winding in the component wave functions. In \cite{PhysRevA.110.L061303}, they emerge from the spin-twist instability of a ferrodark soliton. The anomalous vortices are different from the polar core vortices, which appears in a driven polar condensate, possessing phase winding in the spin components \cite{PhysRevA.108.023308}. For the Q2D condensate, we analyze the radial spin-spin correlations, 
\begin{equation}
C_{\alpha}(\rho)=\frac{1}{N_{\alpha}}dt \int d\rho'\int_0^{2\pi}d\phi F_{\alpha}({\bm \rho}+{\bm \rho}', t)F_{\alpha}({\bm \rho}', t),
\label{cor2}
\end{equation}
where ${\bm \rho}=(x, y)$ is the radial vector with magnitude $\rho=\sqrt{x^2+y^2}$, $\alpha\in\{x, y, z\}$ and $N_{\alpha}=\int d\rho\int d\phi F_{\alpha}({\bm \rho})^2$ for the below-gap case are shown in Fig.~\ref{fig:9}. Unlike in the case of Q1D, the dynamics in Q2D is less sensitive to the initial noise, hence, the time average is not necessary. As expected, $C_{x, y}\sim 0$ [see Fig.~\ref{fig:9}(a)] whereas $C_{z}$ displays a Bessel function dependence [see Fig.~\ref{fig:9}(b)], and in particular we found $C_{z}\sim J_0\left(k_u^{(-)}\rho\right)$. Similar Bessel function correlations are also found in a Q2D driven spinor condensates starting from the polar phase \cite{PhysRevA.108.023308}. When driven above the gap, both longitudinal and transverse spin-spin correlations exhibited the Bessel function behavior, featured by different momenta, i.e, $C_{z}\sim J_0\left(k_u^{(-)}\rho\right)$ and $C_{x}\sim J_0\left(k_u^{(0)}\rho\right)$.

\section{Dynamics under $a_2$ modulation}
\label{a2m}

The Mathieu equations for $a_2$ modulation become 
\begin{eqnarray}
		\frac{d^2 u_{+1}}{dt^2}+\frac{1}{\hbar^2}\left[\epsilon_{+1}(k)^2+\frac{8\alpha_2\bar g_2n_HE_k}{3}\cos(2\omega_2t)\right]u_{+1} = 0,\label{mat1} \\
		\frac{d^2 u_{0}}{dt^2}+\frac{1}{\hbar^2}\left[\epsilon_{0}(k)^2+\frac{4\alpha_2\bar g_2n_H(E_k-q)}{3}\cos(2\omega_2t)\right]u_{0} = 0,\label{mat2} \\
		\frac{d^2 u_{-1}}{dt^2}+\frac{1}{\hbar^2}\left[\epsilon_{-1}(k)^2+\frac{4\alpha_2\bar g_2n_HE_k}{3}\cos(2\omega_2t)\right]u_{-1} = 0. \label{mat3},
	\end{eqnarray}
and the corresponding Floquet exponents are finally obtained as, 
 \begin{eqnarray}
 \label{s20}
 \sigma^{(-)}=\sigma^{(0)}=\frac{\alpha_2n_H\bar g_2}{3\hbar^2\omega_2}\left[\sqrt{(n_H\bar c_1)^2+\hbar^2\omega_2^2}-n_H\bar c_1\right],\\
 \sigma^{(+)}=\frac{2\alpha_2 n_H\bar g_2}{3\hbar^2\omega_2}\left[\sqrt{(n_H\bar c_0)^2+\hbar^2\omega_2^2}-n_H\bar c_0\right].
 \label{s21}
 \end{eqnarray}
Note that Eq.~(\ref{s20}) is identical to Eq.~(\ref{s0}), whereas Eq.~(\ref{s21}) has an additional factor of 2 compared to Eq.~(\ref{s1}). The latter implies that modulating $a_2$ can result in different dynamics than modulating $a_0$ for a given set of system parameters. For instance, when $\bar c_0=\bar c_1$, above the gap, modulation of $a_0$ leads to the emergence of spin and density Faraday patterns as well as spin dynamics. In contrast, with $a_2$ modulation, since $\sigma^{(-)}, \sigma^{(0)}<\sigma^{(+)}$, the spin dynamics hardly occurs, and also the density pattern prevails over the spin pattern, regardless of the driving frequency. With $a_2$ modulation, to induce spin dynamics, we require sufficiently small $\bar c_1$, but not zero and to be more precise, $0<\bar c_1n_H < \left(5 \bar c_0 n_H- 3\sqrt{n_H^2\bar c_0^2 + \hbar^2\omega_2^2}\right)/4$. Apart from that, no contrasting dynamical scenarios emerge while modulating $a_2$ compared to $a_0$ modulation.

\section{Dynamics under $a_0$ and $a_2$ modulation: Competing instabilities}
\label{a02m}

\begin{figure}
\centering
\includegraphics[width= .95\columnwidth]{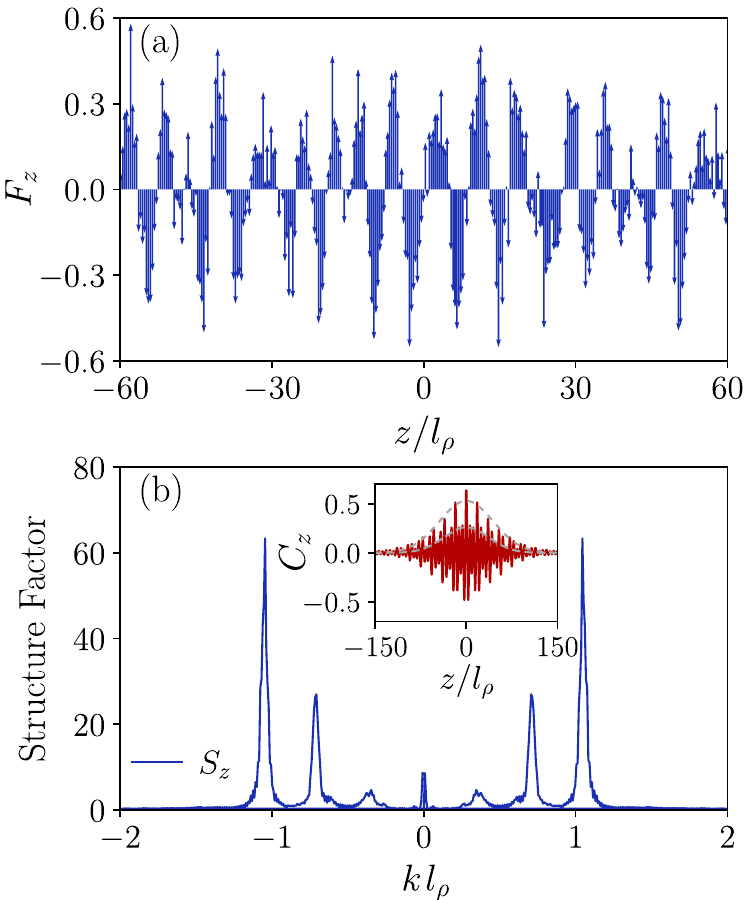}
\caption{\small{(color online). Results for simultaneous modulation of $a_0$ and $a_2$ with $\bar c_1\ll\bar c_0$ and $\omega_{0, 2}<\Delta$. The values of the parameters are $\bar c_{0}n_H^{1D}/2\pi l_\rho^2\hbar\omega_\rho=0.4$, $\bar c_{1}n_H^{1D}/2\pi l_\rho^2\hbar\omega_\rho=0.05$, $q/\hbar\omega_\rho=-0.6$, $\alpha_0=0.2$, $\omega_0=0.6\omega_\rho$, $\alpha_2=0.15$ and $\omega_2=0.3\omega_\rho$, which give a gap of $\Delta=0.648\omega_\rho$. (a) Spin texture for simultaneous modulation of $a_0$ and $a_2$ at $\omega_{\rho} t=268$. The corresponding $C_z$ correlation is shown in the inset of (b). The correlation is obtained by taking an average from $\omega_\rho t_i=180$ to $\omega_\rho t_f=270$ over 40 different realizations of TWA noise. (b) The structure factor revealing the two dominant unstable momenta of the magnon mode.}}
\label{fig:10} 
\end{figure}

In this section, we analyze the dynamics that occur under the simultaneous modulation of $a_0$ and $a_2$, particularly focusing on the intriguing scenario of competing instability when modulated at different frequencies ($\omega_0\neq \omega_2$) and appropriately chosen amplitudes ($\alpha_0\neq \alpha_2$). In the competing instability scenario discussed here, two different momenta of the same Bogoliubov branch become equally unstable and compete in the dynamics. It contrasts with the dynamics discussed in Sec .~\ref {atg}, where three different momenta become equally unstable, each belonging to different Bogoliubov branches. When considering the modulation frequencies below the gap, for $\bar c_0\ll\bar c_1$ and with an appropriate choice of modulation amplitudes, the Mathieu exponent $\sigma^{(+)}$ in Eqs.~(\ref{s21}) ($a_0$-modulation) and (\ref{s1}) ($a_2$-modulation) can be made equal. It leads to the competing instability of two modes from the lowest lying (density) branch, $\epsilon_{+1}(k)$. The resulting density dynamics is then characterized by an oscillation between the patterns of two different wavelengths, identical to that observed for an initial polar phase \cite{PhysRevA.108.023308}. On the other hand, for $\bar c_1\ll\bar c_0$, the two modes associated with the magnon excitation branch $\epsilon_{-1}(k)$ become unstable, resulting in a spin Faraday pattern or spin texture shown in Fig.~\ref{fig:10}(a), unique to the AFM phase. The resulting spin-spin correlation $C_z$ [inset of  Fig.~\ref{fig:10}(b)] has oscillations corresponding to both wavelengths, which is revealed by the two dominant peaks in the structure factor (the Fourier transform of $C_z$), as shown in Fig.~\ref{fig:10}(b). $C_z$ also resembles two Gaussian envelops embedded on each other, as shown by dashed lines.

\begin{figure*}
\centering
\includegraphics[width= 2.\columnwidth]{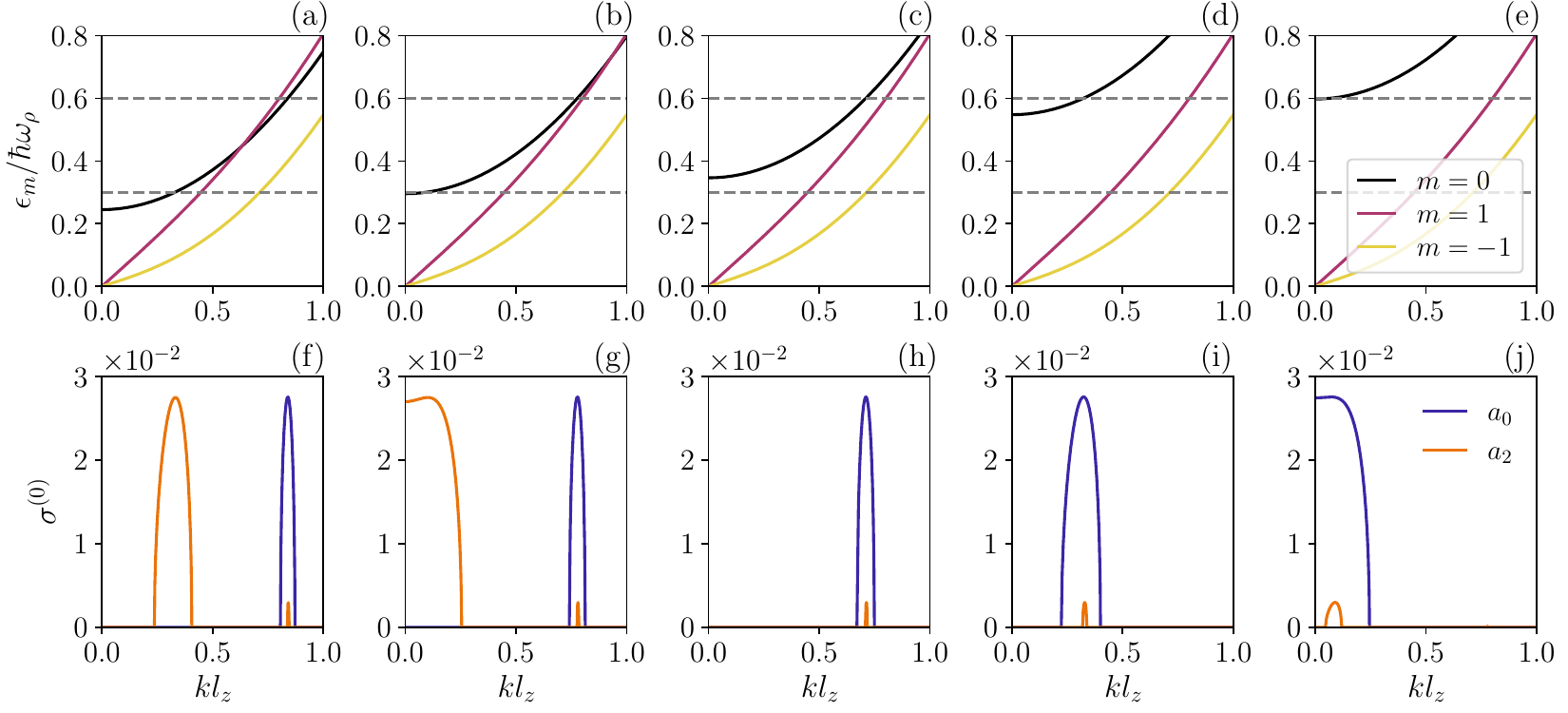}
\caption{\small{(color online). (a)-(e) The Bogoliubov modes ($\epsilon_m$) of a Q1D condensate in AFM phase for $\bar c_{0}n_H^{1D}/2\pi l_\rho^2\hbar\omega_\rho=0.4$, $\bar c_{1}n_H^{1D}/2\pi l_\rho^2\hbar\omega_\rho=0.05$ (a) $q/\hbar\omega_\rho=-0.2$, (b) -0.25, (c) -0.3, (d) -0.5 and (e) -0.55. The modulation frequencies  $\omega_0/\omega_\rho=0.6$, $\omega_2/\omega_\rho=0.3$ are indicated by dashed lines. (f)-(j) The Mathieu exponents associated with the unstable spin-mode shown (a)-(e) for modulation amplitudes, $\alpha_0=0.3$ and $\alpha_2=0.217$. The second blue peak arises from the $a_0$-modulationa and the first orange one arising from the $a_2$ modulation.}}
\label{fig:11} 
\end{figure*}
\begin{figure}
\centering
\includegraphics[width= 1.\columnwidth]{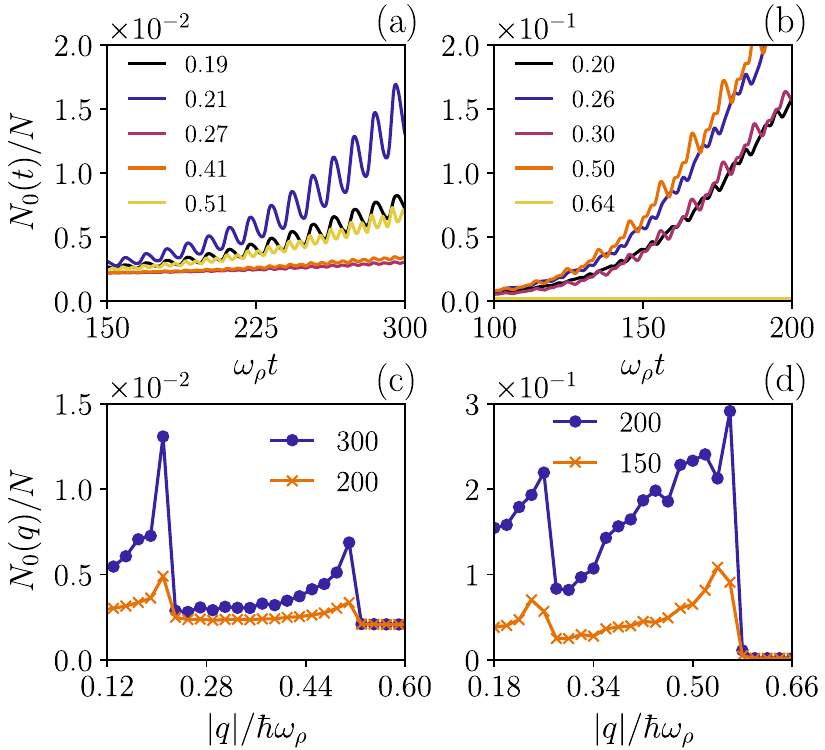}
\caption{\small{(color online). (a) and (b) shows the population dynamics for different values of $|q|$ (given in the inset in units of $\hbar\omega_\rho$) given in the inset for $\bar c_1=\bar c_0$ and  $\bar c_1\ll\bar c_0$, respectively. (c) and (d) show the total population transferred to the $m=0$ component after two different instants (given in the inset in units of $\omega_\rho$) for $\bar c_1=\bar c_0$ and  $\bar c_1\ll\bar c_0$, respectively. For (a) and (c), we took $\bar c_{0}n_H^{1D}/2\pi l_\rho^2\hbar\omega_\rho=\bar c_{1}n_H^{1D}/2\pi l_\rho^2\hbar\omega_\rho=0.1$,  $\alpha_0=0.4$ and $\alpha_2=0.235$.  For (b) and (d), we took $\bar c_{0}n_H^{1D}/2\pi l_\rho^2\hbar\omega_\rho=0.4$, $\bar c_{1}n_H^{1D}/2\pi l_\rho^2\hbar\omega_\rho=0.05$, $\alpha_0=0.3$ and $\alpha_2=0.217$. For all plots, $\omega_0/\omega_\rho=0.6$ and $\omega_2/\omega_\rho=0.3$.}}
\label{fig:12} 
\end{figure}

The two additional scenarios involving competing instability are as follows: one can select a modulation frequency above the gap and the other below it, or both frequencies can be situated above the gap. Strikingly, since the Mathieu exponents are independent of $q$ and the gap ($\Delta$) of the spin mode  $\epsilon_{0}(k)$ can be tuned by $q$, these two cases, including the above case of both frequencies lying below the gap, can be achieved by simply varying $q$. It is demonstrated in Figs.~\ref{fig:11}(a)-\ref{fig:11}(e) using the Bogoliubov spectrum of a Q1D condensate for $\bar c_1\ll\bar c_0$. The dashed horizontal lines in Figs.~\ref{fig:11}(a)-\ref{fig:11}(e) show the modulation frequencies with the upper one for the $a_0$-modulation and the lower one for the $a_2$-modulation. Note that the gapped spin mode $\epsilon_{0}(k)$ shifts upwards in energy as $|q|$ increases. In Fig.~\ref{fig:11}(a), both modulation frequencies lie above the gap, and as $|q|$ increases [see Figs.~\ref{fig:11}(b) and \ref{fig:11}(c)], the modulation frequency $\omega_2$ now appear below the gap. Further increment in $|q|$ leads to the scenario where both frequencies ($\omega_0$ and $\omega_2$) lie below the gap. As we will focus on the population transfer due to the unstable spin mode $\epsilon_0(k)$, we first examine the corresponding Mathieu exponent as a function of the quasi-momentum $k$. They are shown in Figs.~\ref{fig:11}(f)-\ref{fig:11}(j), associated with the spectrum given in Figs.~\ref{fig:11}(a)-\ref{fig:11}(e), respectively. The identical peak values of the primary instability tongues arising from $a_0$ (second peak) and $a_2$ (first peak) modulations shown in Fig.~\ref{fig:11}(f) indicate the competing instability. Since $\bar c_1\ll\bar c_0$, the momenta associated with the magnon mode also identically contribute to the instability dynamics, which govern the spin Faraday pattern wavelengths. As $|q|$ increases, the first peak shifts to $k=0$, broadens, and eventually disappears when $\omega_2$ falls below the gap [see Figs.~\ref{fig:11}(f) and \ref{fig:11}(g) ]. The same happens for the second peak as well upon increasing $|q|$ further, as shown in Figs.~\ref{fig:11}(h)-\ref{fig:11}(j).

The above behavior of instability tongues as a function of $|q|$ leads to a non-trivial variation in the total population transferred from $m=\pm 1$ components to $m=0$ after a fixed time [see Fig.~\ref{fig:12}], despite the peak values of the Mathieu exponents being hardly affected by the change in $|q|$. In particular, Figs.~\ref{fig:12}(a) and \ref{fig:12}(b) reveal a non-monotonous behavior in the dynamics of $N_0(t)$ (the population in the $m=0$ component) for different $|q|$, shown respectively for $\bar c_1\ll\bar c_0$ and $\bar c_1=\bar c_0$. Since a larger $\bar c_1$ makes the spin mode harder, the population transfer is slower in the case of $\bar c_1=\bar c_0$. In Figs.~\ref{fig:12}(c) and \ref{fig:12}(d), we show the population at two different instants for each case as a function of $|q|$, revealing two abrupt jumps. These jumps, which occur from a high population to a lower one as $|q|$ increases, mark the value of $|q|$ at which driving frequencies ($\omega_0$ and $\omega_2$) fall below the spin mode. The increase in the population transfer between the jumps is due to the broadening of the instability tongues due to weaker dependence of spin mode on $k$ as $k\to 0$. This broadening allows more quasi-momenta from the spin mode to contribute to the population transfer. No population transfer occurs for $|q|$ values after the second jump, as both $\omega_0$ and $\omega_2$ lie below the spin mode. Even though, it is not shown, the same behavior can also be observed in a Q2D condensate.

\section{Summary and Outlook}
\label{sok}
In summary, we analyzed the dynamics of spin-1 Q1D and Q2D homogeneous condensates under the modulation of two $s$-wave scattering lengths, starting from an antiferromagnetic phase. In the case of single modulation, driving the $a_0$ scattering length captures all the scenarios. Further we analyzed the interesting scenario of competing instability by simultaneously modulating both scattering lengths with distinct driving frequencies and modulation amplitudes. Dimensionality of the condensate played a vital role in determining the nature of spin textures and spin-spin correlations. In Q2D, we observed anomalous vortices that exhibited no phase winding in their magnetic components. In contrast, Q1D features spin textures characterized by domain walls or random spin orientations, depending on the driving frequency below or above the gapped mode. Interestingly, the nature of spin-spin correlations changed drastically from Q1D to Q2D. As we have shown, in Q2D, the spatial spin-spin correlations exhibited a Bessel function behaviour, whereas in Q1D, an unusual Gaussian envelope is found. The simultaneous modulation of two scattering lengths resulted in a competing instability that affected the population transfer in a complex manner when the QZF was varied.

Our studies open up several perspectives for future studies. One interesting problem is to analyze how the correlation function and spin texture behave at the cross-over between Q1D and Q2D. This requires an analysis that involves calculating Bogoliubov modes in the presence of a harmonic potential along one axis, which can be adjusted to tune the condensate from Q1D to Q2D regimes. Additionally, it would be interesting to explore the same behavior in three-dimensional condensates, investigating condensates with higher spins, for instance spin-2 or spin-3 condensates, and also extending the analysis beyond the linear regime \cite{PhysRevA.109.L051301}.

\section{Acknowledgements}
We thank National Supercomputing Mission (NSM) for providing computing resources of "PARAM Brahma" at IISER Pune, which is implemented by C-DAC and supported by the Ministry of Electronics and Information Technology (MeitY) and Department of Science and Technology (DST), Government of India. We further acknowledge DST-SERB for Swarnajayanti fellowship File No. SB/SJF/2020-21/19, and the MATRICS grant (MTR/2022/000454) from SERB, Government of India and National Mission on Interdisciplinary Cyber-Physical Systems (NM-ICPS) of the Department of Science and Technology, Government of India, through the I-HUB Quantum Technology Foundation, Pune, India.

\bibliography{spin1.bib}

\begin{thebibliography}{43}%
\makeatletter
\providecommand \@ifxundefined [1]{%
 \@ifx{#1\undefined}
}%
\providecommand \@ifnum [1]{%
 \ifnum #1\expandafter \@firstoftwo
 \else \expandafter \@secondoftwo
 \fi
}%
\providecommand \@ifx [1]{%
 \ifx #1\expandafter \@firstoftwo
 \else \expandafter \@secondoftwo
 \fi
}%
\providecommand \natexlab [1]{#1}%
\providecommand \enquote  [1]{``#1''}%
\providecommand \bibnamefont  [1]{#1}%
\providecommand \bibfnamefont [1]{#1}%
\providecommand \citenamefont [1]{#1}%
\providecommand \href@noop [0]{\@secondoftwo}%
\providecommand \href [0]{\begingroup \@sanitize@url \@href}%
\providecommand \@href[1]{\@@startlink{#1}\@@href}%
\providecommand \@@href[1]{\endgroup#1\@@endlink}%
\providecommand \@sanitize@url [0]{\catcode `\\12\catcode `\$12\catcode
  `\&12\catcode `\#12\catcode `\^12\catcode `\_12\catcode `\%12\relax}%
\providecommand \@@startlink[1]{}%
\providecommand \@@endlink[0]{}%
\providecommand \url  [0]{\begingroup\@sanitize@url \@url }%
\providecommand \@url [1]{\endgroup\@href {#1}{\urlprefix }}%
\providecommand \urlprefix  [0]{URL }%
\providecommand \Eprint [0]{\href }%
\providecommand \doibase [0]{https://doi.org/}%
\providecommand \selectlanguage [0]{\@gobble}%
\providecommand \bibinfo  [0]{\@secondoftwo}%
\providecommand \bibfield  [0]{\@secondoftwo}%
\providecommand \translation [1]{[#1]}%
\providecommand \BibitemOpen [0]{}%
\providecommand \bibitemStop [0]{}%
\providecommand \bibitemNoStop [0]{.\EOS\space}%
\providecommand \EOS [0]{\spacefactor3000\relax}%
\providecommand \BibitemShut  [1]{\csname bibitem#1\endcsname}%
\let\auto@bib@innerbib\@empty
\bibitem [{\citenamefont {Jose}\ \emph {et~al.}(2023)\citenamefont {Jose},
  \citenamefont {Sah},\ and\ \citenamefont {Nath}}]{PhysRevA.108.023308}%
  \BibitemOpen
  \bibfield  {author} {\bibinfo {author} {\bibfnamefont {S.~M.}\ \bibnamefont
  {Jose}}, \bibinfo {author} {\bibfnamefont {K.}~\bibnamefont {Sah}},\ and\
  \bibinfo {author} {\bibfnamefont {R.}~\bibnamefont {Nath}},\ }\bibfield
  {title} {\bibinfo {title} {Patterns, spin-spin correlations, and competing
  instabilities in driven quasi-two-dimensional spin-1 bose-einstein
  condensates},\ }\href {https://doi.org/10.1103/PhysRevA.108.023308}
  {\bibfield  {journal} {\bibinfo  {journal} {Phys. Rev. A}\ }\textbf {\bibinfo
  {volume} {108}},\ \bibinfo {pages} {023308} (\bibinfo {year}
  {2023})}\BibitemShut {NoStop}%
\bibitem [{\citenamefont {Cominotti}\ \emph {et~al.}(2022)\citenamefont
  {Cominotti}, \citenamefont {Berti}, \citenamefont {Farolfi}, \citenamefont
  {Zenesini}, \citenamefont {Lamporesi}, \citenamefont {Carusotto},
  \citenamefont {Recati},\ and\ \citenamefont
  {Ferrari}}]{PhysRevLett.128.210401}%
  \BibitemOpen
  \bibfield  {author} {\bibinfo {author} {\bibfnamefont {R.}~\bibnamefont
  {Cominotti}}, \bibinfo {author} {\bibfnamefont {A.}~\bibnamefont {Berti}},
  \bibinfo {author} {\bibfnamefont {A.}~\bibnamefont {Farolfi}}, \bibinfo
  {author} {\bibfnamefont {A.}~\bibnamefont {Zenesini}}, \bibinfo {author}
  {\bibfnamefont {G.}~\bibnamefont {Lamporesi}}, \bibinfo {author}
  {\bibfnamefont {I.}~\bibnamefont {Carusotto}}, \bibinfo {author}
  {\bibfnamefont {A.}~\bibnamefont {Recati}},\ and\ \bibinfo {author}
  {\bibfnamefont {G.}~\bibnamefont {Ferrari}},\ }\bibfield  {title} {\bibinfo
  {title} {Observation of massless and massive collective excitations with
  faraday patterns in a two-component superfluid},\ }\href
  {https://doi.org/10.1103/PhysRevLett.128.210401} {\bibfield  {journal}
  {\bibinfo  {journal} {Phys. Rev. Lett.}\ }\textbf {\bibinfo {volume} {128}},\
  \bibinfo {pages} {210401} (\bibinfo {year} {2022})}\BibitemShut {NoStop}%
\bibitem [{\citenamefont {Hoang}\ \emph {et~al.}(2013)\citenamefont {Hoang},
  \citenamefont {Gerving}, \citenamefont {Land}, \citenamefont {Anquez},
  \citenamefont {Hamley},\ and\ \citenamefont {Chapman}}]{hoa13}%
  \BibitemOpen
  \bibfield  {author} {\bibinfo {author} {\bibfnamefont {T.~M.}\ \bibnamefont
  {Hoang}}, \bibinfo {author} {\bibfnamefont {C.~S.}\ \bibnamefont {Gerving}},
  \bibinfo {author} {\bibfnamefont {B.~J.}\ \bibnamefont {Land}}, \bibinfo
  {author} {\bibfnamefont {M.}~\bibnamefont {Anquez}}, \bibinfo {author}
  {\bibfnamefont {C.~D.}\ \bibnamefont {Hamley}},\ and\ \bibinfo {author}
  {\bibfnamefont {M.~S.}\ \bibnamefont {Chapman}},\ }\bibfield  {title}
  {\bibinfo {title} {Dynamic stabilization of a quantum many-body spin
  system},\ }\href {https://doi.org/10.1103/PhysRevLett.111.090403} {\bibfield
  {journal} {\bibinfo  {journal} {Phys. Rev. Lett.}\ }\textbf {\bibinfo
  {volume} {111}},\ \bibinfo {pages} {090403} (\bibinfo {year}
  {2013})}\BibitemShut {NoStop}%
\bibitem [{\citenamefont {Saito}\ and\ \citenamefont {Hyuga}(2008)}]{sai08}%
  \BibitemOpen
  \bibfield  {author} {\bibinfo {author} {\bibfnamefont {H.}~\bibnamefont
  {Saito}}\ and\ \bibinfo {author} {\bibfnamefont {H.}~\bibnamefont {Hyuga}},\
  }\bibfield  {title} {\bibinfo {title} {Dynamical casimir effect for magnons
  in a spinor bose-einstein condensate},\ }\href
  {https://doi.org/10.1103/PhysRevA.78.033605} {\bibfield  {journal} {\bibinfo
  {journal} {Phys. Rev. A}\ }\textbf {\bibinfo {volume} {78}},\ \bibinfo
  {pages} {033605} (\bibinfo {year} {2008})}\BibitemShut {NoStop}%
\bibitem [{\citenamefont {Hoang}\ \emph {et~al.}(2016)\citenamefont {Hoang},
  \citenamefont {Anquez}, \citenamefont {Robbins}, \citenamefont {Yang},
  \citenamefont {Land}, \citenamefont {Hamley},\ and\ \citenamefont
  {Chapman}}]{hoa16}%
  \BibitemOpen
  \bibfield  {author} {\bibinfo {author} {\bibfnamefont {T.~M.}\ \bibnamefont
  {Hoang}}, \bibinfo {author} {\bibfnamefont {M.}~\bibnamefont {Anquez}},
  \bibinfo {author} {\bibfnamefont {B.~A.}\ \bibnamefont {Robbins}}, \bibinfo
  {author} {\bibfnamefont {X.~Y.}\ \bibnamefont {Yang}}, \bibinfo {author}
  {\bibfnamefont {B.~J.}\ \bibnamefont {Land}}, \bibinfo {author}
  {\bibfnamefont {C.~D.}\ \bibnamefont {Hamley}},\ and\ \bibinfo {author}
  {\bibfnamefont {M.~S.}\ \bibnamefont {Chapman}},\ }\bibfield  {title}
  {\bibinfo {title} {Parametric excitation and squeezing in a many-body spinor
  condensate},\ }\href {https://doi.org/10.1038/ncomms11233} {\bibfield
  {journal} {\bibinfo  {journal} {Nat. Commun.}\ }\textbf {\bibinfo {volume}
  {7}},\ \bibinfo {pages} {11233} (\bibinfo {year} {2016})}\BibitemShut
  {NoStop}%
\bibitem [{\citenamefont {Evrard}\ \emph {et~al.}(2019)\citenamefont {Evrard},
  \citenamefont {Qu}, \citenamefont {Jim\'enez-Garc\'{\i}a}, \citenamefont
  {Dalibard},\ and\ \citenamefont {Gerbier}}]{evr19}%
  \BibitemOpen
  \bibfield  {author} {\bibinfo {author} {\bibfnamefont {B.}~\bibnamefont
  {Evrard}}, \bibinfo {author} {\bibfnamefont {A.}~\bibnamefont {Qu}}, \bibinfo
  {author} {\bibfnamefont {K.}~\bibnamefont {Jim\'enez-Garc\'{\i}a}}, \bibinfo
  {author} {\bibfnamefont {J.}~\bibnamefont {Dalibard}},\ and\ \bibinfo
  {author} {\bibfnamefont {F.}~\bibnamefont {Gerbier}},\ }\bibfield  {title}
  {\bibinfo {title} {Relaxation and hysteresis near shapiro resonances in a
  driven spinor condensate},\ }\href
  {https://doi.org/10.1103/PhysRevA.100.023604} {\bibfield  {journal} {\bibinfo
   {journal} {Phys. Rev. A}\ }\textbf {\bibinfo {volume} {100}},\ \bibinfo
  {pages} {023604} (\bibinfo {year} {2019})}\BibitemShut {NoStop}%
\bibitem [{\citenamefont {Imaeda}\ \emph {et~al.}(2021)\citenamefont {Imaeda},
  \citenamefont {Fujimoto},\ and\ \citenamefont {Kawaguchi}}]{ima21}%
  \BibitemOpen
  \bibfield  {author} {\bibinfo {author} {\bibfnamefont {Y.}~\bibnamefont
  {Imaeda}}, \bibinfo {author} {\bibfnamefont {K.}~\bibnamefont {Fujimoto}},\
  and\ \bibinfo {author} {\bibfnamefont {Y.}~\bibnamefont {Kawaguchi}},\
  }\bibfield  {title} {\bibinfo {title} {Spin-wave growth via shapiro
  resonances in a spinor bose-einstein condensate},\ }\href
  {https://doi.org/10.1103/PhysRevResearch.3.043090} {\bibfield  {journal}
  {\bibinfo  {journal} {Phys. Rev. Research}\ }\textbf {\bibinfo {volume}
  {3}},\ \bibinfo {pages} {043090} (\bibinfo {year} {2021})}\BibitemShut
  {NoStop}%
\bibitem [{\citenamefont {Xu}\ and\ \citenamefont {Zhang}(2021)}]{pen21}%
  \BibitemOpen
  \bibfield  {author} {\bibinfo {author} {\bibfnamefont {P.}~\bibnamefont
  {Xu}}\ and\ \bibinfo {author} {\bibfnamefont {W.}~\bibnamefont {Zhang}},\
  }\bibfield  {title} {\bibinfo {title} {Generalized parametric resonance in a
  spin-1 bose-einstein condensate},\ }\href
  {https://doi.org/10.1103/PhysRevA.104.023324} {\bibfield  {journal} {\bibinfo
   {journal} {Phys. Rev. A}\ }\textbf {\bibinfo {volume} {104}},\ \bibinfo
  {pages} {023324} (\bibinfo {year} {2021})}\BibitemShut {NoStop}%
\bibitem [{\citenamefont {Cheng}(2010)}]{PhysRevA.81.023619}%
  \BibitemOpen
  \bibfield  {author} {\bibinfo {author} {\bibfnamefont {J.}~\bibnamefont
  {Cheng}},\ }\bibfield  {title} {\bibinfo {title} {Chaotic dynamics in a
  periodically driven spin-1 condensate},\ }\href
  {https://doi.org/10.1103/PhysRevA.81.023619} {\bibfield  {journal} {\bibinfo
  {journal} {Phys. Rev. A}\ }\textbf {\bibinfo {volume} {81}},\ \bibinfo
  {pages} {023619} (\bibinfo {year} {2010})}\BibitemShut {NoStop}%
\bibitem [{\citenamefont {Kim}\ \emph {et~al.}(2024)\citenamefont {Kim},
  \citenamefont {Jung}, \citenamefont {Lee}, \citenamefont {Hong},\ and\
  \citenamefont {Shin}}]{PhysRevResearch.6.L032030}%
  \BibitemOpen
  \bibfield  {author} {\bibinfo {author} {\bibfnamefont {J.}~\bibnamefont
  {Kim}}, \bibinfo {author} {\bibfnamefont {J.}~\bibnamefont {Jung}}, \bibinfo
  {author} {\bibfnamefont {J.}~\bibnamefont {Lee}}, \bibinfo {author}
  {\bibfnamefont {D.}~\bibnamefont {Hong}},\ and\ \bibinfo {author}
  {\bibfnamefont {Y.}~\bibnamefont {Shin}},\ }\bibfield  {title} {\bibinfo
  {title} {Chaos-assisted turbulence in spinor bose-einstein condensates},\
  }\href {https://doi.org/10.1103/PhysRevResearch.6.L032030} {\bibfield
  {journal} {\bibinfo  {journal} {Phys. Rev. Res.}\ }\textbf {\bibinfo {volume}
  {6}},\ \bibinfo {pages} {L032030} (\bibinfo {year} {2024})}\BibitemShut
  {NoStop}%
\bibitem [{\citenamefont {Liu}\ \emph {et~al.}(2022)\citenamefont {Liu},
  \citenamefont {Meng}, \citenamefont {Qin},\ and\ \citenamefont
  {Zhou}}]{LIU2022106091}%
  \BibitemOpen
  \bibfield  {author} {\bibinfo {author} {\bibfnamefont {C.-J.}\ \bibnamefont
  {Liu}}, \bibinfo {author} {\bibfnamefont {Y.-C.}\ \bibnamefont {Meng}},
  \bibinfo {author} {\bibfnamefont {J.-L.}\ \bibnamefont {Qin}},\ and\ \bibinfo
  {author} {\bibfnamefont {L.}~\bibnamefont {Zhou}},\ }\bibfield  {title}
  {\bibinfo {title} {Classical and quantum chaos in a spin-1 atomic
  bose–einstein condensate via floquet driving},\ }\href
  {https://doi.org/https://doi.org/10.1016/j.rinp.2022.106091} {\bibfield
  {journal} {\bibinfo  {journal} {Results in Physics}\ }\textbf {\bibinfo
  {volume} {43}},\ \bibinfo {pages} {106091} (\bibinfo {year}
  {2022})}\BibitemShut {NoStop}%
\bibitem [{\citenamefont {Hong}\ \emph {et~al.}(2023)\citenamefont {Hong},
  \citenamefont {Lee}, \citenamefont {Kim}, \citenamefont {Jung}, \citenamefont
  {Lee}, \citenamefont {Kang},\ and\ \citenamefont
  {Shin}}]{PhysRevA.108.013318}%
  \BibitemOpen
  \bibfield  {author} {\bibinfo {author} {\bibfnamefont {D.}~\bibnamefont
  {Hong}}, \bibinfo {author} {\bibfnamefont {J.}~\bibnamefont {Lee}}, \bibinfo
  {author} {\bibfnamefont {J.}~\bibnamefont {Kim}}, \bibinfo {author}
  {\bibfnamefont {J.~H.}\ \bibnamefont {Jung}}, \bibinfo {author}
  {\bibfnamefont {K.}~\bibnamefont {Lee}}, \bibinfo {author} {\bibfnamefont
  {S.}~\bibnamefont {Kang}},\ and\ \bibinfo {author} {\bibfnamefont
  {Y.}~\bibnamefont {Shin}},\ }\bibfield  {title} {\bibinfo {title}
  {Spin-driven stationary turbulence in spinor bose-einstein condensates},\
  }\href {https://doi.org/10.1103/PhysRevA.108.013318} {\bibfield  {journal}
  {\bibinfo  {journal} {Phys. Rev. A}\ }\textbf {\bibinfo {volume} {108}},\
  \bibinfo {pages} {013318} (\bibinfo {year} {2023})}\BibitemShut {NoStop}%
\bibitem [{\citenamefont {Jung}\ \emph {et~al.}(2023)\citenamefont {Jung},
  \citenamefont {Lee}, \citenamefont {Kim},\ and\ \citenamefont
  {Shin}}]{PhysRevA.108.043309}%
  \BibitemOpen
  \bibfield  {author} {\bibinfo {author} {\bibfnamefont {J.~H.}\ \bibnamefont
  {Jung}}, \bibinfo {author} {\bibfnamefont {J.}~\bibnamefont {Lee}}, \bibinfo
  {author} {\bibfnamefont {J.}~\bibnamefont {Kim}},\ and\ \bibinfo {author}
  {\bibfnamefont {Y.}~\bibnamefont {Shin}},\ }\bibfield  {title} {\bibinfo
  {title} {Random spin textures in turbulent spinor bose-einstein
  condensates},\ }\href {https://doi.org/10.1103/PhysRevA.108.043309}
  {\bibfield  {journal} {\bibinfo  {journal} {Phys. Rev. A}\ }\textbf {\bibinfo
  {volume} {108}},\ \bibinfo {pages} {043309} (\bibinfo {year}
  {2023})}\BibitemShut {NoStop}%
\bibitem [{\citenamefont {Dadras}\ \emph {et~al.}(2018)\citenamefont {Dadras},
  \citenamefont {Gresch}, \citenamefont {Groiseau}, \citenamefont {Wimberger},\
  and\ \citenamefont {Summy}}]{dad18}%
  \BibitemOpen
  \bibfield  {author} {\bibinfo {author} {\bibfnamefont {S.}~\bibnamefont
  {Dadras}}, \bibinfo {author} {\bibfnamefont {A.}~\bibnamefont {Gresch}},
  \bibinfo {author} {\bibfnamefont {C.}~\bibnamefont {Groiseau}}, \bibinfo
  {author} {\bibfnamefont {S.}~\bibnamefont {Wimberger}},\ and\ \bibinfo
  {author} {\bibfnamefont {G.~S.}\ \bibnamefont {Summy}},\ }\bibfield  {title}
  {\bibinfo {title} {Quantum walk in momentum space with a bose-einstein
  condensate},\ }\href {https://doi.org/10.1103/PhysRevLett.121.070402}
  {\bibfield  {journal} {\bibinfo  {journal} {Phys. Rev. Lett.}\ }\textbf
  {\bibinfo {volume} {121}},\ \bibinfo {pages} {070402} (\bibinfo {year}
  {2018})}\BibitemShut {NoStop}%
\bibitem [{\citenamefont {Dadras}\ \emph {et~al.}(2019)\citenamefont {Dadras},
  \citenamefont {Gresch}, \citenamefont {Groiseau}, \citenamefont {Wimberger},\
  and\ \citenamefont {Summy}}]{dad19}%
  \BibitemOpen
  \bibfield  {author} {\bibinfo {author} {\bibfnamefont {S.}~\bibnamefont
  {Dadras}}, \bibinfo {author} {\bibfnamefont {A.}~\bibnamefont {Gresch}},
  \bibinfo {author} {\bibfnamefont {C.}~\bibnamefont {Groiseau}}, \bibinfo
  {author} {\bibfnamefont {S.}~\bibnamefont {Wimberger}},\ and\ \bibinfo
  {author} {\bibfnamefont {G.~S.}\ \bibnamefont {Summy}},\ }\bibfield  {title}
  {\bibinfo {title} {Experimental realization of a momentum-space quantum
  walk},\ }\href {https://doi.org/10.1103/PhysRevA.99.043617} {\bibfield
  {journal} {\bibinfo  {journal} {Phys. Rev. A}\ }\textbf {\bibinfo {volume}
  {99}},\ \bibinfo {pages} {043617} (\bibinfo {year} {2019})}\BibitemShut
  {NoStop}%
\bibitem [{\citenamefont {Austin-Harris}\ \emph {et~al.}(2024)\citenamefont
  {Austin-Harris}, \citenamefont {Hardesty-Shaw}, \citenamefont {Guan},
  \citenamefont {Binegar}, \citenamefont {Blume}, \citenamefont {Lewis-Swan},\
  and\ \citenamefont {Liu}}]{PhysRevA.109.043309}%
  \BibitemOpen
  \bibfield  {author} {\bibinfo {author} {\bibfnamefont {J.~O.}\ \bibnamefont
  {Austin-Harris}}, \bibinfo {author} {\bibfnamefont {Z.~N.}\ \bibnamefont
  {Hardesty-Shaw}}, \bibinfo {author} {\bibfnamefont {Q.}~\bibnamefont {Guan}},
  \bibinfo {author} {\bibfnamefont {C.}~\bibnamefont {Binegar}}, \bibinfo
  {author} {\bibfnamefont {D.}~\bibnamefont {Blume}}, \bibinfo {author}
  {\bibfnamefont {R.~J.}\ \bibnamefont {Lewis-Swan}},\ and\ \bibinfo {author}
  {\bibfnamefont {Y.}~\bibnamefont {Liu}},\ }\bibfield  {title} {\bibinfo
  {title} {Engineering dynamical phase diagrams with driven lattices in spinor
  gases},\ }\href {https://doi.org/10.1103/PhysRevA.109.043309} {\bibfield
  {journal} {\bibinfo  {journal} {Phys. Rev. A}\ }\textbf {\bibinfo {volume}
  {109}},\ \bibinfo {pages} {043309} (\bibinfo {year} {2024})}\BibitemShut
  {NoStop}%
\bibitem [{\citenamefont {Engels}\ \emph {et~al.}(2007)\citenamefont {Engels},
  \citenamefont {Atherton},\ and\ \citenamefont
  {Hoefer}}]{PhysRevLett.98.095301}%
  \BibitemOpen
  \bibfield  {author} {\bibinfo {author} {\bibfnamefont {P.}~\bibnamefont
  {Engels}}, \bibinfo {author} {\bibfnamefont {C.}~\bibnamefont {Atherton}},\
  and\ \bibinfo {author} {\bibfnamefont {M.~A.}\ \bibnamefont {Hoefer}},\
  }\bibfield  {title} {\bibinfo {title} {Observation of faraday waves in a
  bose-einstein condensate},\ }\href
  {https://doi.org/10.1103/PhysRevLett.98.095301} {\bibfield  {journal}
  {\bibinfo  {journal} {Phys. Rev. Lett.}\ }\textbf {\bibinfo {volume} {98}},\
  \bibinfo {pages} {095301} (\bibinfo {year} {2007})}\BibitemShut {NoStop}%
\bibitem [{\citenamefont {Nguyen}\ \emph {et~al.}(2019)\citenamefont {Nguyen},
  \citenamefont {Tsatsos}, \citenamefont {Luo}, \citenamefont {Lode},
  \citenamefont {Telles}, \citenamefont {Bagnato},\ and\ \citenamefont
  {Hulet}}]{PhysRevX.9.011052}%
  \BibitemOpen
  \bibfield  {author} {\bibinfo {author} {\bibfnamefont {J.~H.~V.}\
  \bibnamefont {Nguyen}}, \bibinfo {author} {\bibfnamefont {M.~C.}\
  \bibnamefont {Tsatsos}}, \bibinfo {author} {\bibfnamefont {D.}~\bibnamefont
  {Luo}}, \bibinfo {author} {\bibfnamefont {A.~U.~J.}\ \bibnamefont {Lode}},
  \bibinfo {author} {\bibfnamefont {G.~D.}\ \bibnamefont {Telles}}, \bibinfo
  {author} {\bibfnamefont {V.~S.}\ \bibnamefont {Bagnato}},\ and\ \bibinfo
  {author} {\bibfnamefont {R.~G.}\ \bibnamefont {Hulet}},\ }\bibfield  {title}
  {\bibinfo {title} {Parametric excitation of a bose-einstein condensate: From
  faraday waves to granulation},\ }\href
  {https://doi.org/10.1103/PhysRevX.9.011052} {\bibfield  {journal} {\bibinfo
  {journal} {Phys. Rev. X}\ }\textbf {\bibinfo {volume} {9}},\ \bibinfo {pages}
  {011052} (\bibinfo {year} {2019})}\BibitemShut {NoStop}%
\bibitem [{\citenamefont {Zhang}\ \emph {et~al.}(2020)\citenamefont {Zhang},
  \citenamefont {Yao}, \citenamefont {Feng}, \citenamefont {Hu},\ and\
  \citenamefont {Chin}}]{zhang20}%
  \BibitemOpen
  \bibfield  {author} {\bibinfo {author} {\bibfnamefont {Z.}~\bibnamefont
  {Zhang}}, \bibinfo {author} {\bibfnamefont {K.-X.}\ \bibnamefont {Yao}},
  \bibinfo {author} {\bibfnamefont {L.}~\bibnamefont {Feng}}, \bibinfo {author}
  {\bibfnamefont {J.}~\bibnamefont {Hu}},\ and\ \bibinfo {author}
  {\bibfnamefont {C.}~\bibnamefont {Chin}},\ }\bibfield  {title} {\bibinfo
  {title} {Pattern formation in a driven bose--einstein condensate},\ }\href
  {https://doi.org/10.1038/s41567-020-0839-3} {\bibfield  {journal} {\bibinfo
  {journal} {Nature Physics}\ }\textbf {\bibinfo {volume} {16}},\ \bibinfo
  {pages} {652} (\bibinfo {year} {2020})}\BibitemShut {NoStop}%
\bibitem [{\citenamefont {Kwon}\ \emph {et~al.}(2021)\citenamefont {Kwon},
  \citenamefont {Mukherjee}, \citenamefont {Huh}, \citenamefont {Kim},
  \citenamefont {Mistakidis}, \citenamefont {Maity}, \citenamefont
  {Kevrekidis}, \citenamefont {Majumder}, \citenamefont {Schmelcher},\ and\
  \citenamefont {Choi}}]{PhysRevLett.127.113001}%
  \BibitemOpen
  \bibfield  {author} {\bibinfo {author} {\bibfnamefont {K.}~\bibnamefont
  {Kwon}}, \bibinfo {author} {\bibfnamefont {K.}~\bibnamefont {Mukherjee}},
  \bibinfo {author} {\bibfnamefont {S.~J.}\ \bibnamefont {Huh}}, \bibinfo
  {author} {\bibfnamefont {K.}~\bibnamefont {Kim}}, \bibinfo {author}
  {\bibfnamefont {S.~I.}\ \bibnamefont {Mistakidis}}, \bibinfo {author}
  {\bibfnamefont {D.~K.}\ \bibnamefont {Maity}}, \bibinfo {author}
  {\bibfnamefont {P.~G.}\ \bibnamefont {Kevrekidis}}, \bibinfo {author}
  {\bibfnamefont {S.}~\bibnamefont {Majumder}}, \bibinfo {author}
  {\bibfnamefont {P.}~\bibnamefont {Schmelcher}},\ and\ \bibinfo {author}
  {\bibfnamefont {J.-y.}\ \bibnamefont {Choi}},\ }\bibfield  {title} {\bibinfo
  {title} {Spontaneous formation of star-shaped surface patterns in a driven
  bose-einstein condensate},\ }\href
  {https://doi.org/10.1103/PhysRevLett.127.113001} {\bibfield  {journal}
  {\bibinfo  {journal} {Phys. Rev. Lett.}\ }\textbf {\bibinfo {volume} {127}},\
  \bibinfo {pages} {113001} (\bibinfo {year} {2021})}\BibitemShut {NoStop}%
\bibitem [{\citenamefont {Liebster}\ \emph {et~al.}(2025)\citenamefont
  {Liebster}, \citenamefont {Sparn}, \citenamefont {Kath}, \citenamefont
  {Duchene}, \citenamefont {Fujii}, \citenamefont {G\"orlitz}, \citenamefont
  {Enss}, \citenamefont {Strobel},\ and\ \citenamefont
  {Oberthaler}}]{PhysRevX.15.011026}%
  \BibitemOpen
  \bibfield  {author} {\bibinfo {author} {\bibfnamefont {N.}~\bibnamefont
  {Liebster}}, \bibinfo {author} {\bibfnamefont {M.}~\bibnamefont {Sparn}},
  \bibinfo {author} {\bibfnamefont {E.}~\bibnamefont {Kath}}, \bibinfo {author}
  {\bibfnamefont {J.}~\bibnamefont {Duchene}}, \bibinfo {author} {\bibfnamefont
  {K.}~\bibnamefont {Fujii}}, \bibinfo {author} {\bibfnamefont {S.~L.}\
  \bibnamefont {G\"orlitz}}, \bibinfo {author} {\bibfnamefont {T.}~\bibnamefont
  {Enss}}, \bibinfo {author} {\bibfnamefont {H.}~\bibnamefont {Strobel}},\ and\
  \bibinfo {author} {\bibfnamefont {M.~K.}\ \bibnamefont {Oberthaler}},\
  }\bibfield  {title} {\bibinfo {title} {Observation of pattern stabilization
  in a driven superfluid},\ }\href {https://doi.org/10.1103/PhysRevX.15.011026}
  {\bibfield  {journal} {\bibinfo  {journal} {Phys. Rev. X}\ }\textbf {\bibinfo
  {volume} {15}},\ \bibinfo {pages} {011026} (\bibinfo {year}
  {2025})}\BibitemShut {NoStop}%
\bibitem [{\citenamefont {Fu}\ \emph {et~al.}(2018)\citenamefont {Fu},
  \citenamefont {Feng}, \citenamefont {Anderson}, \citenamefont {Clark},
  \citenamefont {Hu}, \citenamefont {Andrade}, \citenamefont {Chin},\ and\
  \citenamefont {Levin}}]{PhysRevLett.121.243001}%
  \BibitemOpen
  \bibfield  {author} {\bibinfo {author} {\bibfnamefont {H.}~\bibnamefont
  {Fu}}, \bibinfo {author} {\bibfnamefont {L.}~\bibnamefont {Feng}}, \bibinfo
  {author} {\bibfnamefont {B.~M.}\ \bibnamefont {Anderson}}, \bibinfo {author}
  {\bibfnamefont {L.~W.}\ \bibnamefont {Clark}}, \bibinfo {author}
  {\bibfnamefont {J.}~\bibnamefont {Hu}}, \bibinfo {author} {\bibfnamefont
  {J.~W.}\ \bibnamefont {Andrade}}, \bibinfo {author} {\bibfnamefont
  {C.}~\bibnamefont {Chin}},\ and\ \bibinfo {author} {\bibfnamefont
  {K.}~\bibnamefont {Levin}},\ }\bibfield  {title} {\bibinfo {title} {Density
  waves and jet emission asymmetry in bose fireworks},\ }\href
  {https://doi.org/10.1103/PhysRevLett.121.243001} {\bibfield  {journal}
  {\bibinfo  {journal} {Phys. Rev. Lett.}\ }\textbf {\bibinfo {volume} {121}},\
  \bibinfo {pages} {243001} (\bibinfo {year} {2018})}\BibitemShut {NoStop}%
\bibitem [{\citenamefont {Fujimoto}\ and\ \citenamefont
  {Uchino}(2019)}]{PhysRevResearch.1.033132}%
  \BibitemOpen
  \bibfield  {author} {\bibinfo {author} {\bibfnamefont {K.}~\bibnamefont
  {Fujimoto}}\ and\ \bibinfo {author} {\bibfnamefont {S.}~\bibnamefont
  {Uchino}},\ }\bibfield  {title} {\bibinfo {title} {Floquet spinor bose
  gases},\ }\href {https://doi.org/10.1103/PhysRevResearch.1.033132} {\bibfield
   {journal} {\bibinfo  {journal} {Phys. Rev. Res.}\ }\textbf {\bibinfo
  {volume} {1}},\ \bibinfo {pages} {033132} (\bibinfo {year}
  {2019})}\BibitemShut {NoStop}%
\bibitem [{\citenamefont {Sadler}\ \emph {et~al.}(2006)\citenamefont {Sadler},
  \citenamefont {Higbie}, \citenamefont {Leslie}, \citenamefont
  {Vengalattore},\ and\ \citenamefont {Stamper-Kurn}}]{sad06}%
  \BibitemOpen
  \bibfield  {author} {\bibinfo {author} {\bibfnamefont {L.~E.}\ \bibnamefont
  {Sadler}}, \bibinfo {author} {\bibfnamefont {J.~M.}\ \bibnamefont {Higbie}},
  \bibinfo {author} {\bibfnamefont {S.~R.}\ \bibnamefont {Leslie}}, \bibinfo
  {author} {\bibfnamefont {M.}~\bibnamefont {Vengalattore}},\ and\ \bibinfo
  {author} {\bibfnamefont {D.~M.}\ \bibnamefont {Stamper-Kurn}},\ }\bibfield
  {title} {\bibinfo {title} {Spontaneous symmetry breaking in a quenched
  ferromagnetic spinor bose--einstein condensate},\ }\href
  {https://doi.org/10.1038/nature05094} {\bibfield  {journal} {\bibinfo
  {journal} {Nature}\ }\textbf {\bibinfo {volume} {443}},\ \bibinfo {pages}
  {312 EP } (\bibinfo {year} {2006})}\BibitemShut {NoStop}%
\bibitem [{\citenamefont {Stamper-Kurn}\ and\ \citenamefont
  {Ueda}(2013)}]{sta13}%
  \BibitemOpen
  \bibfield  {author} {\bibinfo {author} {\bibfnamefont {D.~M.}\ \bibnamefont
  {Stamper-Kurn}}\ and\ \bibinfo {author} {\bibfnamefont {M.}~\bibnamefont
  {Ueda}},\ }\bibfield  {title} {\bibinfo {title} {Spinor bose gases:
  Symmetries, magnetism, and quantum dynamics},\ }\href
  {https://doi.org/10.1103/RevModPhys.85.1191} {\bibfield  {journal} {\bibinfo
  {journal} {Rev. Mod. Phys.}\ }\textbf {\bibinfo {volume} {85}},\ \bibinfo
  {pages} {1191} (\bibinfo {year} {2013})}\BibitemShut {NoStop}%
\bibitem [{\citenamefont {Kawaguchi}\ and\ \citenamefont {Ueda}(2012)}]{kaw12}%
  \BibitemOpen
  \bibfield  {author} {\bibinfo {author} {\bibfnamefont {Y.}~\bibnamefont
  {Kawaguchi}}\ and\ \bibinfo {author} {\bibfnamefont {M.}~\bibnamefont
  {Ueda}},\ }\bibfield  {title} {\bibinfo {title} {Spinor bose--einstein
  condensates},\ }\href
  {https://doi.org/https://doi.org/10.1016/j.physrep.2012.07.005} {\bibfield
  {journal} {\bibinfo  {journal} {Phys. Rep.}\ }\textbf {\bibinfo {volume}
  {520}},\ \bibinfo {pages} {253} (\bibinfo {year} {2012})}\BibitemShut
  {NoStop}%
\bibitem [{\citenamefont {Zhang}\ \emph {et~al.}(2009)\citenamefont {Zhang},
  \citenamefont {Naidon},\ and\ \citenamefont {Ueda}}]{zha09}%
  \BibitemOpen
  \bibfield  {author} {\bibinfo {author} {\bibfnamefont {P.}~\bibnamefont
  {Zhang}}, \bibinfo {author} {\bibfnamefont {P.}~\bibnamefont {Naidon}},\ and\
  \bibinfo {author} {\bibfnamefont {M.}~\bibnamefont {Ueda}},\ }\bibfield
  {title} {\bibinfo {title} {Independent control of scattering lengths in
  multicomponent quantum gases},\ }\href
  {https://doi.org/10.1103/PhysRevLett.103.133202} {\bibfield  {journal}
  {\bibinfo  {journal} {Phys. Rev. Lett.}\ }\textbf {\bibinfo {volume} {103}},\
  \bibinfo {pages} {133202} (\bibinfo {year} {2009})}\BibitemShut {NoStop}%
\bibitem [{\citenamefont {Staliunas}\ \emph {et~al.}(2002)\citenamefont
  {Staliunas}, \citenamefont {Longhi},\ and\ \citenamefont
  {de~Valc\'arcel}}]{sta02}%
  \BibitemOpen
  \bibfield  {author} {\bibinfo {author} {\bibfnamefont {K.}~\bibnamefont
  {Staliunas}}, \bibinfo {author} {\bibfnamefont {S.}~\bibnamefont {Longhi}},\
  and\ \bibinfo {author} {\bibfnamefont {G.~J.}\ \bibnamefont
  {de~Valc\'arcel}},\ }\bibfield  {title} {\bibinfo {title} {Faraday patterns
  in bose-einstein condensates},\ }\href
  {https://doi.org/10.1103/PhysRevLett.89.210406} {\bibfield  {journal}
  {\bibinfo  {journal} {Phys. Rev. Lett.}\ }\textbf {\bibinfo {volume} {89}},\
  \bibinfo {pages} {210406} (\bibinfo {year} {2002})}\BibitemShut {NoStop}%
\bibitem [{\citenamefont {Nath}\ and\ \citenamefont
  {Santos}(2010)}]{PhysRevA.81.033626}%
  \BibitemOpen
  \bibfield  {author} {\bibinfo {author} {\bibfnamefont {R.}~\bibnamefont
  {Nath}}\ and\ \bibinfo {author} {\bibfnamefont {L.}~\bibnamefont {Santos}},\
  }\bibfield  {title} {\bibinfo {title} {Faraday patterns in two-dimensional
  dipolar bose-einstein condensates},\ }\href
  {https://doi.org/10.1103/PhysRevA.81.033626} {\bibfield  {journal} {\bibinfo
  {journal} {Phys. Rev. A}\ }\textbf {\bibinfo {volume} {81}},\ \bibinfo
  {pages} {033626} (\bibinfo {year} {2010})}\BibitemShut {NoStop}%
\bibitem [{\citenamefont {Nicolin}\ \emph {et~al.}(2007)\citenamefont
  {Nicolin}, \citenamefont {Carretero-Gonz\'alez},\ and\ \citenamefont
  {Kevrekidis}}]{PhysRevA.76.063609}%
  \BibitemOpen
  \bibfield  {author} {\bibinfo {author} {\bibfnamefont {A.~I.}\ \bibnamefont
  {Nicolin}}, \bibinfo {author} {\bibfnamefont {R.}~\bibnamefont
  {Carretero-Gonz\'alez}},\ and\ \bibinfo {author} {\bibfnamefont {P.~G.}\
  \bibnamefont {Kevrekidis}},\ }\bibfield  {title} {\bibinfo {title} {Faraday
  waves in bose-einstein condensates},\ }\href
  {https://doi.org/10.1103/PhysRevA.76.063609} {\bibfield  {journal} {\bibinfo
  {journal} {Phys. Rev. A}\ }\textbf {\bibinfo {volume} {76}},\ \bibinfo
  {pages} {063609} (\bibinfo {year} {2007})}\BibitemShut {NoStop}%
\bibitem [{\citenamefont {Zhang}\ \emph {et~al.}(2022)\citenamefont {Zhang},
  \citenamefont {Liu},\ and\ \citenamefont {Zhang}}]{PhysRevA.105.063319}%
  \BibitemOpen
  \bibfield  {author} {\bibinfo {author} {\bibfnamefont {H.}~\bibnamefont
  {Zhang}}, \bibinfo {author} {\bibfnamefont {S.}~\bibnamefont {Liu}},\ and\
  \bibinfo {author} {\bibfnamefont {Y.-S.}\ \bibnamefont {Zhang}},\ }\bibfield
  {title} {\bibinfo {title} {Faraday patterns in spin-orbit-coupled
  bose-einstein condensates},\ }\href
  {https://doi.org/10.1103/PhysRevA.105.063319} {\bibfield  {journal} {\bibinfo
   {journal} {Phys. Rev. A}\ }\textbf {\bibinfo {volume} {105}},\ \bibinfo
  {pages} {063319} (\bibinfo {year} {2022})}\BibitemShut {NoStop}%
\bibitem [{\citenamefont {Verma}\ \emph {et~al.}(2017)\citenamefont {Verma},
  \citenamefont {Rapol},\ and\ \citenamefont {Nath}}]{PhysRevA.95.043618}%
  \BibitemOpen
  \bibfield  {author} {\bibinfo {author} {\bibfnamefont {G.}~\bibnamefont
  {Verma}}, \bibinfo {author} {\bibfnamefont {U.~D.}\ \bibnamefont {Rapol}},\
  and\ \bibinfo {author} {\bibfnamefont {R.}~\bibnamefont {Nath}},\ }\bibfield
  {title} {\bibinfo {title} {Generation of dark solitons and their instability
  dynamics in two-dimensional condensates},\ }\href
  {https://doi.org/10.1103/PhysRevA.95.043618} {\bibfield  {journal} {\bibinfo
  {journal} {Phys. Rev. A}\ }\textbf {\bibinfo {volume} {95}},\ \bibinfo
  {pages} {043618} (\bibinfo {year} {2017})}\BibitemShut {NoStop}%
\bibitem [{\citenamefont {\L{}akomy}\ \emph {et~al.}(2012)\citenamefont
  {\L{}akomy}, \citenamefont {Nath},\ and\ \citenamefont
  {Santos}}]{PhysRevA.86.023620}%
  \BibitemOpen
  \bibfield  {author} {\bibinfo {author} {\bibfnamefont {K.}~\bibnamefont
  {\L{}akomy}}, \bibinfo {author} {\bibfnamefont {R.}~\bibnamefont {Nath}},\
  and\ \bibinfo {author} {\bibfnamefont {L.}~\bibnamefont {Santos}},\
  }\bibfield  {title} {\bibinfo {title} {Faraday patterns in coupled
  one-dimensional dipolar condensates},\ }\href
  {https://doi.org/10.1103/PhysRevA.86.023620} {\bibfield  {journal} {\bibinfo
  {journal} {Phys. Rev. A}\ }\textbf {\bibinfo {volume} {86}},\ \bibinfo
  {pages} {023620} (\bibinfo {year} {2012})}\BibitemShut {NoStop}%
\bibitem [{\citenamefont {Nadiger}\ \emph {et~al.}(2024)\citenamefont
  {Nadiger}, \citenamefont {Jose}, \citenamefont {Ghosh}, \citenamefont
  {Kaur},\ and\ \citenamefont {Nath}}]{PhysRevA.109.033309}%
  \BibitemOpen
  \bibfield  {author} {\bibinfo {author} {\bibfnamefont {S.}~\bibnamefont
  {Nadiger}}, \bibinfo {author} {\bibfnamefont {S.~M.}\ \bibnamefont {Jose}},
  \bibinfo {author} {\bibfnamefont {R.}~\bibnamefont {Ghosh}}, \bibinfo
  {author} {\bibfnamefont {I.}~\bibnamefont {Kaur}},\ and\ \bibinfo {author}
  {\bibfnamefont {R.}~\bibnamefont {Nath}},\ }\bibfield  {title} {\bibinfo
  {title} {Stripe and checkerboard patterns in a stack of driven
  quasi-one-dimensional dipolar condensates},\ }\href
  {https://doi.org/10.1103/PhysRevA.109.033309} {\bibfield  {journal} {\bibinfo
   {journal} {Phys. Rev. A}\ }\textbf {\bibinfo {volume} {109}},\ \bibinfo
  {pages} {033309} (\bibinfo {year} {2024})}\BibitemShut {NoStop}%
\bibitem [{\citenamefont {del R\'{\i}o-Lima}\ \emph {et~al.}(2024)\citenamefont
  {del R\'{\i}o-Lima}, \citenamefont {Seman}, \citenamefont {J\'auregui},\ and\
  \citenamefont {Poveda-Cuevas}}]{PhysRevA.110.053318}%
  \BibitemOpen
  \bibfield  {author} {\bibinfo {author} {\bibfnamefont {A.}~\bibnamefont {del
  R\'{\i}o-Lima}}, \bibinfo {author} {\bibfnamefont {J.~A.}\ \bibnamefont
  {Seman}}, \bibinfo {author} {\bibfnamefont {R.}~\bibnamefont {J\'auregui}},\
  and\ \bibinfo {author} {\bibfnamefont {F.~J.}\ \bibnamefont
  {Poveda-Cuevas}},\ }\bibfield  {title} {\bibinfo {title} {Spatial and
  temporal periodic density patterns in driven bose-einstein condensates},\
  }\href {https://doi.org/10.1103/PhysRevA.110.053318} {\bibfield  {journal}
  {\bibinfo  {journal} {Phys. Rev. A}\ }\textbf {\bibinfo {volume} {110}},\
  \bibinfo {pages} {053318} (\bibinfo {year} {2024})}\BibitemShut {NoStop}%
\bibitem [{\citenamefont {Wan}\ \emph {et~al.}(2024)\citenamefont {Wan},
  \citenamefont {Wen},\ and\ \citenamefont {Li}}]{Wan_2024}%
  \BibitemOpen
  \bibfield  {author} {\bibinfo {author} {\bibfnamefont {J.}~\bibnamefont
  {Wan}}, \bibinfo {author} {\bibfnamefont {W.}~\bibnamefont {Wen}},\ and\
  \bibinfo {author} {\bibfnamefont {H.-j.}\ \bibnamefont {Li}},\ }\bibfield
  {title} {\bibinfo {title} {Spatiotemporal pattern formation in parametrically
  driven two-dimensional bose–einstein condensates},\ }\href
  {https://doi.org/10.1088/1572-9494/ad7373} {\bibfield  {journal} {\bibinfo
  {journal} {Communications in Theoretical Physics}\ }\textbf {\bibinfo
  {volume} {76}},\ \bibinfo {pages} {125502} (\bibinfo {year}
  {2024})}\BibitemShut {NoStop}%
\bibitem [{\citenamefont {Jim{\'e}nez-Garc{\'\i}a}\ \emph
  {et~al.}(2019)\citenamefont {Jim{\'e}nez-Garc{\'\i}a}, \citenamefont
  {Invernizzi}, \citenamefont {Evrard}, \citenamefont {Frapolli}, \citenamefont
  {Dalibard},\ and\ \citenamefont {Gerbier}}]{Q1Dspinorexpt}%
  \BibitemOpen
  \bibfield  {author} {\bibinfo {author} {\bibfnamefont {K.}~\bibnamefont
  {Jim{\'e}nez-Garc{\'\i}a}}, \bibinfo {author} {\bibfnamefont
  {A.}~\bibnamefont {Invernizzi}}, \bibinfo {author} {\bibfnamefont
  {B.}~\bibnamefont {Evrard}}, \bibinfo {author} {\bibfnamefont
  {C.}~\bibnamefont {Frapolli}}, \bibinfo {author} {\bibfnamefont
  {J.}~\bibnamefont {Dalibard}},\ and\ \bibinfo {author} {\bibfnamefont
  {F.}~\bibnamefont {Gerbier}},\ }\bibfield  {title} {\bibinfo {title}
  {Spontaneous formation and relaxation of spin domains in antiferromagnetic
  spin-1 condensates},\ }\href {https://doi.org/10.1038/s41467-019-08505-6}
  {\bibfield  {journal} {\bibinfo  {journal} {Nature Communications}\ }\textbf
  {\bibinfo {volume} {10}},\ \bibinfo {pages} {1422} (\bibinfo {year}
  {2019})}\BibitemShut {NoStop}%
\bibitem [{\citenamefont {Kang}\ \emph {et~al.}(2020)\citenamefont {Kang},
  \citenamefont {Hong}, \citenamefont {Kim},\ and\ \citenamefont
  {Shin}}]{PhysRevA.101.023613}%
  \BibitemOpen
  \bibfield  {author} {\bibinfo {author} {\bibfnamefont {S.}~\bibnamefont
  {Kang}}, \bibinfo {author} {\bibfnamefont {D.}~\bibnamefont {Hong}}, \bibinfo
  {author} {\bibfnamefont {J.~H.}\ \bibnamefont {Kim}},\ and\ \bibinfo {author}
  {\bibfnamefont {Y.}~\bibnamefont {Shin}},\ }\bibfield  {title} {\bibinfo
  {title} {Crossover from weak to strong quench in a spinor bose-einstein
  condensate},\ }\href {https://doi.org/10.1103/PhysRevA.101.023613} {\bibfield
   {journal} {\bibinfo  {journal} {Phys. Rev. A}\ }\textbf {\bibinfo {volume}
  {101}},\ \bibinfo {pages} {023613} (\bibinfo {year} {2020})}\BibitemShut
  {NoStop}%
\bibitem [{\citenamefont {Jacob}\ \emph {et~al.}(2012)\citenamefont {Jacob},
  \citenamefont {Shao}, \citenamefont {Corre}, \citenamefont {Zibold},
  \citenamefont {De~Sarlo}, \citenamefont {Mimoun}, \citenamefont {Dalibard},\
  and\ \citenamefont {Gerbier}}]{PhysRevA.86.061601}%
  \BibitemOpen
  \bibfield  {author} {\bibinfo {author} {\bibfnamefont {D.}~\bibnamefont
  {Jacob}}, \bibinfo {author} {\bibfnamefont {L.}~\bibnamefont {Shao}},
  \bibinfo {author} {\bibfnamefont {V.}~\bibnamefont {Corre}}, \bibinfo
  {author} {\bibfnamefont {T.}~\bibnamefont {Zibold}}, \bibinfo {author}
  {\bibfnamefont {L.}~\bibnamefont {De~Sarlo}}, \bibinfo {author}
  {\bibfnamefont {E.}~\bibnamefont {Mimoun}}, \bibinfo {author} {\bibfnamefont
  {J.}~\bibnamefont {Dalibard}},\ and\ \bibinfo {author} {\bibfnamefont
  {F.}~\bibnamefont {Gerbier}},\ }\bibfield  {title} {\bibinfo {title} {Phase
  diagram of spin-1 antiferromagnetic bose-einstein condensates},\ }\href
  {https://doi.org/10.1103/PhysRevA.86.061601} {\bibfield  {journal} {\bibinfo
  {journal} {Phys. Rev. A}\ }\textbf {\bibinfo {volume} {86}},\ \bibinfo
  {pages} {061601} (\bibinfo {year} {2012})}\BibitemShut {NoStop}%
\bibitem [{\citenamefont {Yu}\ and\ \citenamefont
  {Blakie}(2024)}]{PhysRevA.110.L061303}%
  \BibitemOpen
  \bibfield  {author} {\bibinfo {author} {\bibfnamefont {X.}~\bibnamefont
  {Yu}}\ and\ \bibinfo {author} {\bibfnamefont {P.~B.}\ \bibnamefont
  {Blakie}},\ }\bibfield  {title} {\bibinfo {title} {Absence of the breakdown
  of ferrodark solitons exhibiting a snake instability},\ }\href
  {https://doi.org/10.1103/PhysRevA.110.L061303} {\bibfield  {journal}
  {\bibinfo  {journal} {Phys. Rev. A}\ }\textbf {\bibinfo {volume} {110}},\
  \bibinfo {pages} {L061303} (\bibinfo {year} {2024})}\BibitemShut {NoStop}%
\bibitem [{\citenamefont {Stolze}\ \emph {et~al.}(1995)\citenamefont {Stolze},
  \citenamefont {N\"oppert},\ and\ \citenamefont
  {M\"uller}}]{PhysRevB.52.4319}%
  \BibitemOpen
  \bibfield  {author} {\bibinfo {author} {\bibfnamefont {J.}~\bibnamefont
  {Stolze}}, \bibinfo {author} {\bibfnamefont {A.}~\bibnamefont {N\"oppert}},\
  and\ \bibinfo {author} {\bibfnamefont {G.}~\bibnamefont {M\"uller}},\
  }\bibfield  {title} {\bibinfo {title} {Gaussian, exponential, and power-law
  decay of time-dependent correlation functions in quantum spin chains},\
  }\href {https://doi.org/10.1103/PhysRevB.52.4319} {\bibfield  {journal}
  {\bibinfo  {journal} {Phys. Rev. B}\ }\textbf {\bibinfo {volume} {52}},\
  \bibinfo {pages} {4319} (\bibinfo {year} {1995})}\BibitemShut {NoStop}%
\bibitem [{\citenamefont {Bert}\ \emph {et~al.}(2011)\citenamefont {Bert},
  \citenamefont {Kalisky}, \citenamefont {Bell}, \citenamefont {Kim},
  \citenamefont {Hikita}, \citenamefont {Hwang},\ and\ \citenamefont
  {Moler}}]{Bert2011}%
  \BibitemOpen
  \bibfield  {author} {\bibinfo {author} {\bibfnamefont {J.~A.}\ \bibnamefont
  {Bert}}, \bibinfo {author} {\bibfnamefont {B.}~\bibnamefont {Kalisky}},
  \bibinfo {author} {\bibfnamefont {C.}~\bibnamefont {Bell}}, \bibinfo {author}
  {\bibfnamefont {M.}~\bibnamefont {Kim}}, \bibinfo {author} {\bibfnamefont
  {Y.}~\bibnamefont {Hikita}}, \bibinfo {author} {\bibfnamefont {H.~Y.}\
  \bibnamefont {Hwang}},\ and\ \bibinfo {author} {\bibfnamefont {K.~A.}\
  \bibnamefont {Moler}},\ }\bibfield  {title} {\bibinfo {title} {Direct imaging
  of the coexistence of ferromagnetism and superconductivity at the
  laalo3/srtio3 interface},\ }\href {https://doi.org/10.1038/nphys2079}
  {\bibfield  {journal} {\bibinfo  {journal} {Nature Physics}\ }\textbf
  {\bibinfo {volume} {7}},\ \bibinfo {pages} {767} (\bibinfo {year}
  {2011})}\BibitemShut {NoStop}%
\bibitem [{\citenamefont {Fujii}\ \emph {et~al.}(2024)\citenamefont {Fujii},
  \citenamefont {G\"orlitz}, \citenamefont {Liebster}, \citenamefont {Sparn},
  \citenamefont {Kath}, \citenamefont {Strobel}, \citenamefont {Oberthaler},\
  and\ \citenamefont {Enss}}]{PhysRevA.109.L051301}%
  \BibitemOpen
  \bibfield  {author} {\bibinfo {author} {\bibfnamefont {K.}~\bibnamefont
  {Fujii}}, \bibinfo {author} {\bibfnamefont {S.~L.}\ \bibnamefont
  {G\"orlitz}}, \bibinfo {author} {\bibfnamefont {N.}~\bibnamefont {Liebster}},
  \bibinfo {author} {\bibfnamefont {M.}~\bibnamefont {Sparn}}, \bibinfo
  {author} {\bibfnamefont {E.}~\bibnamefont {Kath}}, \bibinfo {author}
  {\bibfnamefont {H.}~\bibnamefont {Strobel}}, \bibinfo {author} {\bibfnamefont
  {M.~K.}\ \bibnamefont {Oberthaler}},\ and\ \bibinfo {author} {\bibfnamefont
  {T.}~\bibnamefont {Enss}},\ }\bibfield  {title} {\bibinfo {title}
  {Stable-fixed-point description of square-pattern formation in driven
  two-dimensional bose-einstein condensates},\ }\href
  {https://doi.org/10.1103/PhysRevA.109.L051301} {\bibfield  {journal}
  {\bibinfo  {journal} {Phys. Rev. A}\ }\textbf {\bibinfo {volume} {109}},\
  \bibinfo {pages} {L051301} (\bibinfo {year} {2024})}\BibitemShut {NoStop}%
\end{thebibliography}%
\end{document}